\documentclass[journal,final]{IEEEtran}
\usepackage{graphics,graphicx,color,float}
\usepackage{amsmath,mathtools}
\usepackage{amssymb}
\usepackage{amsfonts}
\usepackage{amsthm, hyperref}
\usepackage[usenames,dvipsnames]{xcolor}
\usepackage{geometry}
\usepackage{tikz-cd}

\newcommand{\beq}{\begin{equation}}
\newcommand{\enq}{\end{equation}}
\newcommand{\bel}{\begin{lemma}}
\newcommand{\enl}{\end{lemma}}
\newcommand{\bet}{\begin{theorem}}
\newcommand{\ent}{\end{theorem}}

\newcommand{\tr}{\mathrm{Tr}}

\newcommand{\Tr}{\mathrm{Tr}}
\newcommand{\ketbra}[1]{|#1\rangle\langle#1|}

\newcommand{\err}{\mathrm{err}}
\newcommand{\eps}{\varepsilon}

\newcommand{\id}{\ensuremath{\mathrm{I}}}

\newcommand*{\cH}{\mathcal{H}}

\newcommand*{\cD}{\mathcal{D}}

\newcommand*{\cN}{\mathcal{N}}

\newcommand*{\cX}{\mathcal{X}}

\newcommand*{\cE}{\mathcal{E}}

\newcommand{\cP}{\mathcal{P}}

\newcommand{\suppress}[1]{}

\newcommand{\defeq}{\ensuremath{ \stackrel{\mathrm{def}}{=} }}
\newcommand{\F}{\mathrm{F}}
\newcommand{\Pur}{\mathrm{P}}
\newcommand {\br} [1] {\ensuremath{ \left( #1 \right) }}

\newcommand {\minusspace} {\: \! \!}
\newcommand {\smallspace} {\: \!}
\newcommand {\fn} [2] {\ensuremath{ #1 \minusspace \br{ #2 } }}
\newcommand {\ball} [2] {\fn{\mathcal{B}^{#1}}{#2}}
\newcommand {\relent} [2] {\fn{\mathrm{D}}{#1 \middle\| #2}}
\newcommand {\varrelent} [2] {\fn{V}{#1 \middle\| #2}}
\newcommand {\dmax} [2] {\fn{\mathrm{D}_{\max}}{#1 \middle\| #2}}

\newcommand {\dmaxeps} [3] {\fn{\mathrm{D}^{#3}_{\max}}{#1 \middle\| #2}}
\newcommand {\dmineps} [3] {\fn{\mathrm{D}^{#3}_{\mathrm{H}}}{#1 \middle\| #2}}
\newcommand {\dheps} [3] {\fn{\mathrm{D}^{#3}_{\mathrm{H}}}{#1 \middle\| #2}}

\newcommand {\mutinf} [2] {\fn{\mathrm{I}}{#1 \smallspace : \smallspace #2}}
\newcommand {\imax}{\ensuremath{\mathrm{I}_{\max}}}
\newcommand {\imaxeps} [1] {\ensuremath{\mathrm{I}^{#1}_{\max}}}
\newcommand {\condmutinf} [3] {\mutinf{#1}{#2 \smallspace \middle\vert \smallspace #3}}
\newcommand {\condent} [2] {\ensuremath{\mathrm{H}(#1 | #2)}}

\newcommand {\hmax}[2]{\ensuremath{\mathrm{H}_{\max}(#1|#2)}}
\newcommand {\hmin}[2]{\ensuremath{\mathrm{H}_{\min}(#1|#2)}}
\newcommand {\entr}[1]{\ensuremath{\mathrm{S}(#1)}}
\newcommand {\net}{\mathsf{N}}

\newcommand*{\cY}{\mathcal{Y}}

\newcommand*{\cL}{\mathcal{L}}

\newcommand{\bra}[1]{\langle #1|}
\newcommand{\ket}[1]{|#1 \rangle}

\mathchardef\mhyphen="2D

\makeatletter
\newcommand*{\rom}[1]{\expandafter\@slowromancap\romannumeral #1@}
\makeatother

\mathchardef\mhyphen="2D

\newtheorem*{remark}{Remark}
\newtheorem{definition}{Definition}

\newtheorem{fact}{Fact}
\newtheorem{proposition}{Proposition}

\newtheorem{theorem}{Theorem}
\newtheorem{lemma}{Lemma}
\newtheorem{corollary}{Corollary}

\begin {document}
\title{Noisy quantum state redistribution with promise and the Alpha-bit}

\author{Anurag Anshu, Min-Hsiu~Hsieh,~\IEEEmembership{Senior Member,~IEEE,} Rahul Jain
%\thanks{This work was presented in part in ISIT2019.}
\thanks{A. Anshu was affiliated with the Center for Quantum Technologies, National University of Singapore, Singapore, e-mail: (a0109169@u.nus.edu).}
\thanks{Min-Hsiu Hsieh is with the Center for Quantum Software and Information, University of Technology Sydney, Sydney, Australia, e-mail: (min-hsiu.hsieh@uts.edu.au).}
\thanks{R. Jain is with the Center for Quantum Technologies, National University of Singapore, and MajuLab, UMI 3654, Singapore, e-mail: (rahul@comp.nus.edu.sg).}

}

%\author{
%Anurag Anshu\footnote{Center for Quantum Technologies, National University of Singapore, Singapore. \texttt{a0109169@u.nus.edu}} \qquad
%Min-Hsiu Hsieh\footnote{Centre for Quantum Software and Information, University of Technology Sydney, Australia. \texttt{Min-Hsiu.Hsieh@uts.edu.au}} \qquad
%Rahul Jain\footnote{Center for Quantum Technologies, National University of Singapore and MajuLab, UMI 3654, 
%Singapore. \texttt{rahul@comp.nus.edu.sg}} \qquad 
%} 

%\date{}
\maketitle

\begin{abstract}
We consider a variation of the well-studied quantum state redistribution task, in which the starting state is known only to the receiver Bob and not to the sender Alice. We refer to this as quantum state redistribution with a one-sided promise. In addition, we consider communication from Alice to Bob over a noisy channel $\cN$, instead of the noiseless channel, as is usually considered in state redistribution.   We take a natural approach towards the  solution of this problem where we ``embed'' the promise as part of the state and then invoke known protocols for quantum state redistribution composed with known protocols for transfer of quantum information over noisy channels.  
%{We interpret the communication primitive Alpha-bit, recently introduced in Ref.~[\emph{arXiv:1706.09434}], as an instance of state transfer (a sub-task of state redistribution) with a one-sided promise over noisy channels.} 
Using our approach, we are able to reproduce the  Alpha-bit capacities with or without entanglement assistance in Ref.~\cite{1706.09434}, using known protocols for quantum state redistribution and quantum communication over noisy channels. Furthermore, we generalize the entanglement assisted classical Alpha-bit capacity, showing that any quantum state redistribution protocol can be used as a black box to simulate classical communication.
\end{abstract}

\section{Introduction}

{\em Quantum state redistribution}~\cite{Devatakyard, YardD09, YeBW08, Oppenheim14, Berta14, AnshuJW17SR} is a very fundamental and well studied communication task. In this a pure state $\ket{\psi}_{RABC}$ (in the registers $RABC$) is shared between three parties: Reference (R), Alice (AC) and Bob (B).  Alice is supposed to communicate to Bob at the end of which Bob should end up with the register $C$. Alice and Bob may have a state shared between them beforehand and the final state should have high fidelity with the starting state.  The intention is to minimize communication from Alice to Bob and/or minimize entanglement used by the protocol. This task has several applications in quantum network theory and also for {\em direct sum} \cite{Dave14} and {\em direct product} results in quantum communication complexity. Many related sub-tasks have also been studied, namely {\em quantum state merging} (where register $A$ is trivial)~\cite{horodecki07, Berta09, Renner11, ADJ14}, {\em quantum state splitting} (where register $B$ is trivial)~\cite{AbeyesingheDHW09, Renner11, ADJ14} and {\em state transfer} (where registers $A$ and $B$ are trivial)~\cite{Schumacher95}.

In quantum state redistribution, the starting state $\ket{\psi}_{RABC}$ is known to both Alice and Bob. We consider a generalized setting in which the starting state $\ket{\Psi^y}_{RABC}$ is drawn from a (finite) set $\{\ket{\Psi^y}_{RABC}\}_{y \in \cY}$. Alice and Bob both know the set $\cY$; however, $y$ is known only to Bob. We refer to this as quantum state redistribution with (one-sided) {\em promise}. Our motivation for studying this comes from a new communication resource called {\em Alpha-bit} introduced in  Ref.~\cite{1706.09434}. This was introduced in the context of approximate quantum error correction, generalizing the idea of weak decoupling duality introduced in \cite{HaydenW12}. Along with \cite{7530895,KhanianW18, KhanianW19, AnshuLT2019}, this result has made progress on the challenging problem of quantum data compression when the sender has limited classical information about the state to be transmitted. The authors have also shown, in a follow-up work \cite{HaydenP2019}, how the resource of Alpha-bit naturally originates in the quantum information theoretic study of black-holes. 

We interpret the resource of Alpha-bit as an instance of  state transfer with (one-sided) promise. Here the set $\cY$ consists of pure states $\{\ket{\psi}_{RC}^S\}$ which are maximally entangled across $R$ and $C$ and the support of $\psi_R$ and $\psi_C$ is $S$,  which is a subspace of dimension $d^\alpha$ (for some $\alpha > 0$) of the underlying Hilbert space of dimension $d$. Both Alice and Bob know $\cY$; however, only Bob knows $S$. In addition, in the resource framework of Alpha-bit, Alice and Bob are provided with a noisy channel $\cN$ for communication, instead of the noiseless channel provided in usual state transfer. The intention is to maximize $d$, per use of the channel $\cN$. {A related scenario was also studied in \cite{KhanianW18}, with a side-information (a promise, in our terminology) available to Bob about the quantum state shared between Reference and Alice.}

We take a natural approach towards the solution of quantum state redistribution with (one-sided) promise. We ``embed'' the promise inside the state and consider it as a special case of quantum state redistribution itself. We assume the uniform distribution $\mu$ on $\cY$ and consider the following state,  
\begin{equation*}
\ket{\Psi}_{R_YYRABC}\defeq  \sum_{y}\sqrt{\mu(y)}\ket{y,y}_{R_YY}\otimes\ket{\Psi^y}_{RABC}.
\end{equation*}
We consider quantum state redistribution for the state $\ket{\Psi}_{R_YYRABC}$ where the registers $RABC$ are held as usual, the new register $Y$ is held by Bob and the new register $R_Y$ is held by Reference. We then invoke the best known protocols for state redistribution~\cite{Devatakyard, YardD09,  AnshuJW17SR}. 

As in Alpha-bit, we also consider providing a noisy channel $\cN$ between Alice and Bob. In this case we compose the best known protocols for state redistribution with the best known entanglement assisted protocols for transfer of quantum information through noisy channels~\cite{BennettSST02, DevetakHW04, DattaH11, DattaH13, AnshuJW17CC,5550293}. The approach that we take reproduces the achievability bounds on $\alpha$ obtained in Ref.~\cite{1706.09434} in the asymptotic i.i.d setting (with error approaching zero) for both entanglement assisted (Corollary \ref{cor:qachieve}) and unassisted (Theorem \ref{unassistedav}) scenarios for every noisy channel $\cN$. %In other words, we reproduce the entanglement assisted and unassisted Alpha-bit capacities obtained in Ref.~\cite{1706.09434} for every channel $\cN$.  
Ref.~\cite{1706.09434} also considers the scenario where the error needs to be bounded for every subspace $S$ and not just averaged over a uniformly chosen subspace $S$. By considering general distributions over $\cY$ (not just the uniform distribution) and using a minimax theorem we are able to reproduce the bounds obtained in~\cite{1706.09434} in the worst case error setting as well (Corollary~\ref{cor:qachieve} and Theorem~\ref{thm:worstAlpha-bitun}). 
Furthermore, Ref.~\cite[Theorem 5]{1706.09434} shows that the ability to communicate the resource of Alpha-bit (with subspace of dimension $d^\alpha$) provides the ability to transmit $(1+\alpha) \log d$ classical bits with entanglement-assistance. We generalize this to argue that state redistribution protocol for any quantum state $\Psi_{RABC}$ (even mixed) provides the ability to transmit $\condmutinf{R}{C}{B}_{\Psi}$ classical bits, with entanglement-assistance (Theorem~\ref{theo:QSRsim}). As a result, we also recover ~\cite[Theorem 5]{1706.09434}, in Corollary \ref{cor:Alpha-bitsim}.

Finally, we consider a classical version of Alpha-bit (in the presence of noisy channels in Sections~\ref{sec:classicalent} and~\ref{sec:classical}) where the inputs of Alice are drawn from a subset $S$ of size $d^\alpha$ of an underlying set of size $d$. The subset $S$ is known only to Bob. It can be noted that this can be accomplished, with ideal channel between Alice and Bob (as is done by for example by Slepian-Wolf~\cite{SlepianW73}), by Alice sending $\alpha \log d$ random hashes of her input to Bob. This is much better than communication $\frac{2}{1 + \alpha}\log d$ bits  (entanglement assisted) required for Alpha-bit (with ideal channel between Alice and Bob). This can be considered as an evidence against the existence of good ``quantum hashes''.  

%\subsection*{Organization}

We structure our paper as follows. In Section~\ref{sec:prelim}, we introduce notation and definitions of the relevant entropic quantities.  In Section~\ref{sec:qsrp}, we present a collection of one-shot and asymptotic i.i.d. bounds for quantum state redistribution with (one-sided) promise in the presence of noisy channels. In Section~\ref{sec:ptop}, we apply these bounds to recover the Alpha-bit capacities obtained in \cite{1706.09434}. We also consider the classical analogue of Alpha-bit in this section. In Section \ref{sec:qsrresource}, we show how any quantum state redistribution protocol can be used as a resource for entanglement assisted communication of classical messages. 

\section{Preliminaries}
\label{sec:prelim}

Consider a finite dimensional Hilbert space $\cH$ endowed with an inner product $\langle \cdot, \cdot \rangle$ (in this paper, we only consider finite dimensional Hilbert-spaces). The $\ell_1$ norm of an operator $X$ on $\cH$ is $\| X\|_1:=\Tr\sqrt{X^{\dagger}X}$ and $\ell_2$ norm is $\| X\|_2:=\sqrt{\Tr XX^{\dagger}}$. A quantum state (or a density matrix or a state) is a positive semi-definite matrix on $\cH$ with trace equal to $1$. It is called {\em pure} if and only if its rank is $1$. A sub-normalized state is a positive semi-definite matrix on $\cH$ with trace less than or equal to $1$. Let $\ket{\psi}$ be a unit vector on $\cH$, that is $\langle \psi,\psi \rangle=1$.  With some abuse of notation, we use $\psi$ to represent the state and also the density matrix $\ketbra{\psi}$, associated with $\ket{\psi}$. Given a quantum state $\rho$ on $\cH$, the {\em support of $\rho$}, called $\text{supp}(\rho)$ is the subspace of $\cH$ spanned by all eigen-vectors of $\rho$ with non-zero eigenvalues.
 
A {\em quantum register} $A$ is associated with some Hilbert space $\cH_A$. Define $|A| := \dim(\cH_A)$. Let $\mathcal{L}(A)$ represent the set of all linear operators on $\cH_A$. Let $\mathcal{P}(A)$ represent the set of all positive semidefinite operators on $\cH_A$. We denote by $\mathcal{D}(A)$, the set of quantum states on the Hilbert space $\cH_A$. A state $\rho$ with subscript $A$ indicates $\rho_A \in \mathcal{D}(A)$. If two registers $A,B$ are associated with the same Hilbert space, we shall represent the relation by $A\equiv B$.  The composition of two registers $A$ and $B$, denoted $AB$, is associated with the Hilbert space $\cH_A \otimes \cH_B$.  For two quantum states $\rho\in \mathcal{D}(A)$ and $\sigma\in \mathcal{D}(B)$, $\rho\otimes\sigma \in \mathcal{D}(AB)$ represents the tensor product (Kronecker product) of $\rho$ and $\sigma$. The identity operator on $\cH_A$ (and associated register $A$) is denoted $\id_A$. For any operator $O$ on $\cH_A$, we denote by $\{O\}_+$ the subspace spanned by the non-negative eigenvalues of $O$ and by $\{O\}_-$ the subspace spanned by the negative eigenvalues of $O$.  For a positive semidefinite operator $M\in \mathcal{P}(A)$, the largest and smallest non-zero eigenvalues of $M$ are denoted by $\lambda_{\max}(M)$ and $\lambda_{\min}(M)$, respectively.

Let $\rho_{AB} \in \mathcal{D}(AB)$. We define
\[ \rho_{B} := \Tr_{A}\rho_{AB}
:= \sum_i (\bra{i} \otimes \id_{B})
\rho_{AB} (\ket{i} \otimes \id_{B}) , \]
where $\{\ket{i}\}_i$ is an orthonormal basis for the Hilbert space $\cH_A$.
The state $\rho_B\in \mathcal{D}(B)$ is referred to as the marginal state of $\rho_{AB}$. Unless otherwise stated, the missing register from subscript in a state will represent partial trace over that register. Given a $\rho_A\in\mathcal{D}(A)$, a {\em purification} of $\rho_A$ is a pure state $\rho_{AB}\in \mathcal{D}(AB)$ such that $\Tr_{B}{\rho_{AB}}=\rho_A$. A purification of a quantum state is not unique.

A quantum channel $\cE: \mathcal{L}(A)\rightarrow \mathcal{L}(B)$ is a completely positive and trace preserving (CPTP) linear map (mapping states in $\mathcal{D}(A)$ to states in $\mathcal{D}(B)$). A {\em unitary} operator $U_A:\cH_A \rightarrow \cH_A$ is such that $U_A^{\dagger}U_A = U_A U_A^{\dagger} = \id_A$. An {\em isometry}  $V:\cH_A \rightarrow \cH_B$ is such that $V^{\dagger}V = \id_A$. The set of all unitary operators on register $A$ is  denoted by $\mathcal{U}(A)$.

Let $\varepsilon \in (0,1)$. We shall consider the following information theoretic quantities. All the logarithms appearing below are in base $2$. 
\begin{enumerate}
\item {\bf Fidelity} (\cite{uhlmann76}, see also \cite{Josza94}) For $\rho_A,\sigma_A \in \mathcal{D}(A)$, $$\F(\rho_A,\sigma_A)\defeq\|\sqrt{\rho_A}\sqrt{\sigma_A}\|_1.$$ For classical probability distributions $P = \{p_i\}, Q =\{q_i\}$, $$\F(P,Q)\defeq \sum_i \sqrt{p_i \cdot q_i}.$$
\item {\bf Purified distance} (\cite{GilchristLN05, Tomamichel12}) For $\rho_A,\sigma_A \in \mathcal{D}(A)$, $$\Pur(\rho_A,\sigma_A) \defeq \sqrt{1-\F^2(\rho_A,\sigma_A)}.$$

\item {\bf $\varepsilon$-ball} For $\rho_A\in \mathcal{D}(A)$, $$\ball{\eps}{\rho_A} \defeq \{\rho'_A\in \mathcal{D}(A)|~\Pur(\rho_A,\rho'_A) \leq \varepsilon\}. $$ 
\item {\bf Von-Neumann entropy} (\cite{Neumann32}) For $\rho_A\in\mathcal{D}(A)$, $$\entr{\rho_A} \defeq - \Tr(\rho_A\log\rho_A).$$

\item {\bf Relative entropy} (\cite{umegaki1954}) For $\rho_A,\sigma_A\in \mathcal{D}(A)$ such that $\text{supp}(\rho_A) \subseteq \text{supp}(\sigma_A)$, $$\relent{\rho_A}{\sigma_A} \defeq \Tr(\rho_A\log\rho_A) - \Tr(\rho_A\log\sigma_A) .$$ 

\item {\bf Relative entropy variance} For $\rho_A,\sigma_A\in \mathcal{D}(A)$ such that $\text{supp}(\rho_A) \subseteq \text{supp}(\sigma_A)$, $$V(\rho\|\sigma) \defeq \Tr(\rho(\log\rho - \log\sigma)^2) - (\relent{\rho}{\sigma})^2.$$

\item {\bf Mutual information} For $\rho_{AB}\in \mathcal{D}(AB)$, \begin{eqnarray*}
\mutinf{A}{B}_{\rho} &\defeq& \entr{\rho_A} + \entr{\rho_B}-\entr{\rho_{AB}} \\
 &=& \relent{\rho_{AB}}{\rho_A\otimes\rho_B}.
\end{eqnarray*}

\item {\bf Conditional mutual information} For $\rho_{ABC}\in\mathcal{D}(ABC)$, $$\condmutinf{A}{B}{C}_{\rho}\defeq \mutinf{A}{BC}_{\rho}-\mutinf{A}{C}_{\rho}.$$
\item {\bf Conditional entropy}  For $\rho_{AB}\in \mathcal{D}(AB)$, $$\condent{A}{B}_{\rho} \defeq \entr{\rho_{AB}} - \entr{\rho_B}.$$

\item {\bf Max-relative entropy} (\cite{Datta09}) For $\rho_A,\sigma_A\in \mathcal{P}(A)$ such that $\text{supp}(\rho_A) \subseteq \text{supp}(\sigma_A)$, $$ \dmax{\rho_A}{\sigma_A}  \defeq  \inf \{ \lambda \in \mathbb{R} : 2^{\lambda} \sigma_A \succeq \rho_A \}  .$$  
\item {\bf Smooth max-relative entropy} (\cite{Datta09}, see also \cite{Jain:2009}) For $\rho_A\in \mathcal{D}(A) ,\sigma_A\in \mathcal{P}(A)$ such that $\text{supp}(\rho_A) \subseteq \text{supp}(\sigma_A)$, $$ \dmaxeps{\rho_A}{\sigma_A}{\eps}  \defeq  \inf_{\rho'_A\in \ball{\eps}{\rho_A}} \dmax{\rho_A'}{\sigma_A}  .$$  
\item {\bf Smooth hypothesis testing divergence} (\cite{BuscemiD10}, see also \cite{HayashiN03, WangR12}) For $\rho_A\in \mathcal{D}(A) ,\sigma_A\in \mathcal{P}(A)$, $$ \dmineps{\rho_A}{\sigma_A}{\eps}  \defeq  \sup_{0\preceq\Pi\preceq I, \Tr(\Pi\rho_A)\geq 1-\eps}\log\left(\frac{1}{\Tr(\Pi\sigma_A)}\right).$$  

\item {\bf Max-information} (\cite{Renner11}) For $\rho_{AB}\in \mathcal{D}(AB)$, $$ \imax(A:B)_{\rho} \defeq   \inf_{\sigma_{B}\in \mathcal{D}(B)}\dmax{\rho_{AB}}{\rho_{A}\otimes\sigma_{B}} .$$
\item {\bf Smooth max-information} (\cite{Renner11}) For $\rho_{AB}\in \mathcal{D}(AB)$,  $$\imaxeps{\eps}(A:B)_{\rho} \defeq \inf_{\rho'\in \ball{\eps}{\rho}} \imax(A:B)_{\rho'} .$$	

\item {\bf Conditional min-entropy} (\cite{Renner05}) For $\rho_{AB}\in \mathcal{D}(AB)$, define $$ \hmin{A}{B}_{\rho} \defeq  - \inf_{\sigma_B\in \mathcal{D}(B)}\dmax{\rho_{AB}}{\id_{A}\otimes\sigma_{B}} .$$  	
\item {\bf Conditional max-entropy} (\cite{Renner05}) For $\rho_{AB}\in \mathcal{D}(AB)$, define $$\hmax{A}{B}_{\rho} \defeq \max_{\sigma_B\in\mathcal{D}(B)}\log\F^2(\rho_{AB},\id_A\otimes \sigma_B).$$
\end{enumerate}

We will use the following facts. 
\begin{fact}[Triangle inequality for purified distance,~\cite{GilchristLN05, Tomamichel12}]
\label{fact:trianglepurified}
For states $\rho_A, \sigma_A, \tau_A\in \mathcal{D}(A)$,
$$\Pur(\rho_A,\sigma_A) \leq \Pur(\rho_A,\tau_A)  + \Pur(\tau_A,\sigma_A) . $$ 
\end{fact}

\begin{fact}[Monotonicity under quantum channels, \cite{barnum96, lindblad75, Datta09, WangR12}]
	\label{fact:monotonequantumoperation}
For quantum states $\rho$, $\sigma \in \mathcal{D}(A)$, and a quantum channel $\cE(\cdot):\mathcal{L}(A)\rightarrow \mathcal{L}(B)$, it holds that
\begin{align*}
 \dmax{\cE(\rho)}{\cE(\sigma)} &\leq \dmax{\rho}{\sigma}  \\
\F(\cE(\rho),\cE(\sigma)) &\geq \F(\rho,\sigma) \\
 \dmineps{\rho}{\sigma}{\eps} &\geq \dmineps{\cE(\rho)}{\cE(\sigma)}{\eps}.
\end{align*}
In particular, for bipartite states $\rho_{AB},\sigma_{AB}\in \mathcal{D}(AB)$, it holds that
\begin{align*}
\dmax{\rho_{AB}}{\sigma_{AB}} &\geq \dmax{\rho_A}{\sigma_A} \\
\F(\rho_{AB},\sigma_{AB}) &\leq \F(\rho_A,\sigma_A)\\
 \dmineps{\rho_{AB}}{\sigma_{AB}}{\eps} &\geq \dmineps{\rho_A}{\sigma_A}{\eps}.
\end{align*}
\end{fact}

\begin{fact}[Uhlmann's Theorem, \cite{uhlmann76}]
\label{uhlmann}
Let $\rho_A,\sigma_A\in \mathcal{D}(A)$. Let $\rho_{AB}\in \mathcal{D}(AB)$ be a purification of $\rho_A$ and $\ket{\sigma}_{AC}\in\mathcal{D}(AC)$ be a purification of $\sigma_A$. There exists an isometry $V: C \rightarrow B$ such that,
 $$\F(\ketbra{\theta}_{AB}, \ketbra{\rho}_{AB}) = \F(\rho_A,\sigma_A) ,$$
 where $\ket{\theta}_{AB} = (\id_A \otimes V) \ket{\sigma}_{AC}$.
\end{fact}

\begin{fact}[Fannes inequality, \cite{fannes73}]
\label{fact:fannes}
Given quantum states $\rho_1,\rho_2\in \cD(\cH_A)$, such that $|A|=d$ and 
$$\Pur(\rho_1,\rho_2)= \eps \leq \frac{1}{2\mathrm{e}},$$ then
$$|\entr{\rho_1}-\entr{\rho_2}|\leq \eps\log(d)+1.$$   
\end{fact}

\begin{fact}[Hayashi-Nagaoka inequality, \cite{HayashiN03}]
\label{haynag}
Let $0\preceq S\preceq \id,T$ be positive semi-definite operators and $c > 0$. Then 
$$\id - (S+T)^{-\frac{1}{2}}S(S+T)^{-\frac{1}{2}}\preceq (1+c)(\id-S) + \left(2 + c + \frac{1}{c}\right) T.$$
\end{fact}

\begin{fact}[\cite{TomHay13, li2014}]
\label{dmaxequi}
Let $\eps\in (0,1)$ and $n$ be an integer. Let $\rho^{\otimes n}, \sigma^{\otimes n}$ be quantum states. Define 
$$\Phi(x) = \int_{-\infty}^x \frac{e^{-t^2/2}}{\sqrt{2\pi}} dt.$$ It holds that
\begin{equation*}
\dmaxeps{\rho^{\otimes n}}{\sigma^{\otimes n}}{\eps} = n\relent{\rho}{\sigma} - \sqrt{n\varrelent{\rho}{\sigma}} \Phi^{-1}(\eps) + O(\log n) ,
\end{equation*}
and 
\begin{equation*}
\dmineps{\rho^{\otimes n}}{\sigma^{\otimes n}}{\eps} = n\relent{\rho}{\sigma} + \sqrt{n\varrelent{\rho}{\sigma}} \Phi^{-1}(\eps) + O(\log n) .
\end{equation*}
\end{fact}

\begin{fact}
\label{gaussianupper}
For the function $\Phi(x) = \int_{-\infty}^x \frac{e^{-t^2/2}}{\sqrt{2\pi}} dt$ and $\eps\leq \frac{1}{2}$, it holds that $|\Phi^{-1}(\eps)| \leq \sqrt{2\log\frac{1}{2\eps}}$.
\end{fact}
\begin{proof}
We have 
\begin{align*}
\Phi(-x)&=\int_{-\infty}^{-x} \frac{e^{-t^2/2}}{\sqrt{2\pi}} dt \\
&= \int_{0}^{\infty} \frac{e^{-(-x-t)^2/2}}{\sqrt{2\pi}} dt \\
&\leq e^{-x^2/2} \int_{0}^{\infty} \frac{e^{-(-t)^2/2}}{\sqrt{2\pi}} dt \\
&= \frac{1}{2}e^{-x^2/2}.
\end{align*}Thus, $\Phi^{-1}(\eps) \geq -\sqrt{2\log\frac{1}{2\eps}}$. On the other hand, $\Phi^{-1}(\eps)\leq 0$ for $\eps\leq \frac{1}{2}$, which completes the proof.
\end{proof}

%\begin{fact}[\cite{AnshuJW17CC}, see also Appendix B of \cite{AnshuHW20}]
%\label{fact:err}
%Let $\rho$ and $\sigma$ be quantum states and $\Lambda$ be such that $0\preceq \Lambda \preceq \mathbb{I}$. Then
%$$\left|\sqrt{\tr\left[\Lambda\rho\right]}-\sqrt{\tr\left[\Lambda\sigma\right]}\right| \leq \Pur(\rho,\sigma).$$
%\end{fact}

\begin{fact}
\label{fact:err}
Let $\rho$ and $\sigma$ be quantum states and $\Lambda$ be such that $0\preceq \Lambda \preceq \mathbb{I}$. Then
$$\left|\tr\left[\Lambda\rho\right]-\tr\left[\Lambda\sigma\right]\right| \leq \Pur(\rho,\sigma).$$
\end{fact}
\begin{proof}
We have 
$$\left|\tr\left[\Lambda\rho\right]-\tr\left[\Lambda\sigma\right]\right|\leq \frac{1}{2}\|\rho-\sigma\|_1.$$
Since $\frac{1}{2}\|\rho-\sigma\|_1\leq \sqrt{1-\F(\rho,\sigma)}\leq \Pur(\rho,\sigma)$, the proof concludes.
\end{proof}

\begin{fact}[Theorem 5, \cite{AnshuJW17comp}]
\label{fact:Ihdhsame}
Let $\rho_{AB}$ be a quantum state and $\eps \in (0,1)$. For every $\delta>0$, it holds that 
\begin{align*}
\dheps{\rho_{AB}}{\rho_A\otimes \rho_{B}}{\eps -\delta}- 2\log\frac{\eps}{\delta} &\leq \inf_{\sigma_B}\dheps{\rho_{AB}}{\rho_A\otimes \sigma_{B}}{\eps} \\
& \leq \dheps{\rho_{AB}}{\rho_A\otimes \rho_{B}}{\eps}.
\end{align*}
\end{fact}

\begin{fact}
\label{fact:smoothdh}
Let $\eps, \delta\in (0,1)$ and $\rho, \sigma$ be quantum states such that $\Pur(\rho, \sigma)\leq \delta$. Then for any quantum state $\tau$, 
$$\dheps{\rho}{\tau}{\eps+\delta} \geq \dheps{\sigma}{\tau}{\eps}.$$
\end{fact}
\begin{proof}
Let $\Lambda$ be the operator achieving the supremum in the definition of $\dheps{\sigma}{\tau}{\eps}$. Then $\Tr(\Lambda\sigma)\geq 1-\eps$. Invoking Fact \ref{fact:err}, this implies that $\Tr(\Lambda\rho)\geq 1-\eps-\delta$. Further,
$$2^{-\dheps{\sigma}{\tau}{\eps}} = \Tr(\Lambda\sigma) \geq 2^{-\dheps{\rho}{\tau}{\eps+\delta}},$$ by the definition of $\dheps{\rho}{\tau}{\eps+\delta}$. This completes the proof.
\end{proof}

\begin{fact}[Minimax theorem \cite{vonNeumann1928}]
\label{fact:minimax}
Let $\cX,\cY$ be convex compact sets and $f:\cX\times \cY\rightarrow \mathbb{R}$ be a continuous function that satisfies the following properties: $f(\cdot ,y): \cX\rightarrow \mathbb {R}$ is convex for fixed $y$, and
$f(x,\cdot ):\cY\rightarrow \mathbb {R}$  is concave for fixed $x$. Then it holds that
$$\min_{x\in \cX}\max_{y\in \cY} f(x,y) = \max_{y\in \cY}\min_{x\in \cX} f(x,y).$$
\end{fact}

\section{Quantum state redistribution with promise}
\label{sec:qsrp}
In this section, we formally define the communication tasks and present our capacity theorems for them. We begin with a definition of quantum state redistribution, in a slightly general context that also involves mixed states. In the pure state case, and also assuming that the state shared between Alice and Bob is maximally entangled in the beginning and at the end of the protocol, our definition reduces to the standard one \cite{Devatakyard, YardD09}.

\begin{definition}[Quantum state redistribution]
\label{def:qsr}
Fix an $\eps\in (0,1)$, and consider the state $\Psi_{RABC}$. A $(q ,\epsilon)$-quantum state redistribution protocol consists of 
\begin{itemize}
\item an encoding isometry $V: \cH_{ACE_A} \rightarrow \cH_{AQT_A}$  by Alice, and
\item a decoding isometry  $W: \cH_{QBE_B}\rightarrow \cH_{BCT_B}$ by Bob,
\end{itemize}
such that 
$$\Pur\left(\Tr_{T_AT_B}\left(WV(\Psi_{RABC}\otimes \theta_{E_AE_B})V^{\dagger}W^{\dagger}\right),\Psi_{RABC}\right)\leq \eps,$$
where  $\ket{\theta}_{E_AE_B}$ is an entangled state shared between Alice ($E_A$) and Bob ($E_B$).
 The number of qubits communicated is $q = \log|Q|$. 
\end{definition}

Now we introduce the framework considered in our paper. Let $\mathcal{Y}$ be a collection of promises, where each instance $y\in \mathcal{Y}$ occurs with probability $p(y)$. Let $Y$ be the register containing the promises. We assume, throughout the paper, that this register $Y$ is only accessible to Bob, but not to Alice. The goal is for Alice to make as few as possible uses of a noisy channel $\mathcal{N}$ to transmit her quantum system, denoted by $C$. Prior to the communication of a system $C$ over a noisy quantum channel, the state shared between the sender and receiver, that carries the promise $y$, can be viewed as $\ket{\Psi^y}_{RABC}$, where the sender Alice holds $A$ and $C$, Bob holds $B$, while Reference holds $R$. Such a state includes information not only about a promise $y$, but also side information at Alice's side as well as at Bob's side. Moreover, denote 
\begin{equation}\label{eq_generalQCP}
\ket{\Psi}_{R_YYRABC}\defeq  \sum_{y}\sqrt{p(y)}\ket{y,y}_{R_YY}\otimes\ket{\Psi^y}_{RABC},
\end{equation}
with $AC$ belonging to Alice, $BY$ to Bob and $R_YR$ to Reference. 

We formally define a $(n,\epsilon)$-$\overline{\text{QCP}}$ code for sending $C$ with an average error over the channel $\mathcal{N}_{J\to K}$ as follows. Here, we abbreviate \emph{quantum communication with a one-sided promise} as QCP.

\begin{definition}
\label{def:avgcqredist}
%[Average case classical-quantum state redistribution task] 

Fix an $\eps\in (0,1)$.  Consider the state $\ket{\Psi}_{R_YYRABC}$ defined in Eq.~(\ref{eq_generalQCP}), and let Alice ($E_A$) and Bob ($E_B$) pre-share an entangled state $\ket{\theta}_{E_AE_B}$. A $(n,\epsilon)$-$\overline{\rm{QCP}}$ code over the quantum channel $\mathcal{N}_{J\to K}$ with an average error $\eps$ consists of 
\begin{itemize}
\item Alice's encoding map $\cE: \cL(ACE_A) \rightarrow \cL(A J^{\otimes n})$, where the register $J^{\otimes n}$ is communicated with $n$ uses of the channel $\cN_{J\to K}$, and
\item Bob's decoding $\cD: \cL(K^{\otimes n}YB)\rightarrow \cL(YBC)$.  
\end{itemize}
Let the final state be
$$\Phi_{R_YYRABC} \defeq \cD\circ\cN^{\otimes n}\circ\cE(\Psi_{R_YYRABC}\otimes \theta_{E_AE_B}).$$ 
It holds that upon tracing out register $R_Y$,
$$\Pur(\Phi_{YRABC},\Psi_{YRABC})\leq \eps.$$
\end{definition}

In addition, we can also define a $(n,\epsilon)$-${\downarrow}\text{QCP}$ code for sending $C$ with the worst case error over the channel $\mathcal{N}_{J\to K}$ as follows.

\begin{definition}
Fix an $\eps\in (0,1)$. Consider the state $\ket{\Psi}_{R_YYRABC}$ defined in Eq.~(\ref{eq_generalQCP}), and let Alice ($E_A$) and Bob ($E_B$) pre-share an entangled state $\ket{\theta}_{E_AE_B}$. A $(n,\epsilon)$-${\downarrow\rm{QCP}}$ code over the quantum channel $\mathcal{N}_{J\to K}$ with the worst case error consists of 
\begin{itemize}
\item Alice's encoding map $\cE: \cL(ACE_A) \rightarrow \cL(A J^{\otimes n})$ and the register $J^{\otimes n}$ is communicated with $n$ uses of the channel $\cN_{J\to K}$, and  
\item Bob's decoding map $\cD^y: \cL(K^{\otimes n}B)\rightarrow \cL(BC)$.  
\end{itemize}
Let the final state be 
$$\Phi^y_{RABC} \defeq \cD^y\circ\cN^{\otimes n}\circ\cE(\Psi^y_{RABC}\otimes \theta_{E_AE_B}).$$ 
It holds that, $\forall y\in\mathcal{Y}$, 
$$\Pur(\Phi^y_{RABC},\Psi^y_{RABC})\leq \eps.$$
\end{definition}

\begin{remark}[Relationship to quantum state redistribution] Recall the state
$$
\ket{\Psi}_{R_YYRABC}\defeq  \sum_{y}\sqrt{p(y)}\ket{y,y}_{R_YY}\otimes\ket{\Psi^y}_{RABC},
$$
prior to the communication of $C$ from Alice to Bob, with $AC$ belonging to Alice, $BY$ to Bob and $R_YR$ to Reference. For each fixed $y$, in case Alice and Bob both know $y$, the task  is quantum state redistribution over a noisy quantum channel (instead of an error free channel), i.e., to redistribute $C$ subsystem of $\ket{\Psi^y}_{RABC}$ from Alice to Bob. 
\end{remark}

The following result follows from the bounds given in \cite{AnshuJW17SR}.

\begin{proposition}[Achievability bound for ideal qubit channel]
\label{averagecase}
Fix $\eps_1,\eps_2\in (0,1)$. There exists a $(\ell_q, 3\eps_1+5\eps_2)$-quantum state redistribution protocol for the quantum state $\Psi_{YRABC}$.
\end{proposition}
\begin{proof}
We apply the bound given in \cite{AnshuJW17SR} on the quantum state $\Psi_{R_YYRABC}$. The final state $\Phi_{R_YYRABC}$ satisfies 
$$\Pur(\Phi_{R_YYRABC}, \Psi_{R_YYRABC})\leq 3\eps_1+5\eps_2,$$ which implies the desired bound by tracing out register $R_Y$.
\end{proof}

\bigskip
Our first two results are as follows. 

\begin{theorem}[Achievability bound]
\label{averagecasenoise}
Fix $\eps_1,\eps_2, \delta_1,\delta_2 \in (0,1)$ and a quantum channel $\cN_{J \rightarrow K}$. Define $$C^{\delta_1}(\cN_{J \rightarrow K}) \defeq \max_{\ketbra{\theta}_{JJ'}}\dmineps{\cN_{J\rightarrow K}(\theta_{JJ'})}{\cN_{J\rightarrow K}(\theta_{J})\otimes \theta_{J'}}{\delta_1}.$$
 There exists a $(n,3\eps_1+5\eps_2 + 2\sqrt{2\delta_1+2\delta_2})$-$\overline{\rm{QCP}}$ code for the quantum state $\ket{\Psi}_{R_YYRABC}$ with $n$, that is the number of uses of the quantum channel $\cN_{J \rightarrow K}$ between Alice to Bob, upper bounded by 
$$\max\left(\frac{2\ell_q}{C^{\delta_1/ 2\ell_q}(\cN_{J \rightarrow K}) + 2\log\delta_2 - 2\log (2\ell_q)},1\right),$$ where $\ell_q$
 is the minimum of
\begin{multline*}
\frac{1}{2}\inf_{\sigma_C}\left( \inf_{\Psi'\in \ball{\eps_1}{\Psi}}\dmax{\Psi'_{R_YRAC}}{\Psi'_{R_YRA}\otimes \sigma_C}\right. \\  -\biggl. \dmineps{\Psi_{AC}}{\Psi_{A}\otimes \sigma_C}{\eps_2}\biggr) + \log\left(\frac{1}{\varepsilon_1\cdot \eps_2}\right),
\end{multline*}
and 
\begin{multline*}
\frac{1}{2}\inf_{\sigma_C} \left( \inf_{\Psi'\in \ball{\eps_1}{\Psi}}\dmax{\Psi'_{R_YYRBC}}{\Psi'_{R_YYRB}\otimes \sigma_C}\right. \\ -\biggl. \dmineps{\Psi_{YBC}}{\Psi_{YB}\otimes \sigma_C}{\eps_2}\biggr) + \log\left(\frac{1}{\varepsilon_1\cdot \eps_2}\right).
\end{multline*}
\suppress{In other words, there exists an entanglement-assisted one-way protocol which takes as input $\Psi_{R_YYRABC}$ shared between three parties Reference ($R_YR$), Bob ($BY$) and Alice ($AC$) and outputs a state $\Phi_{R_YYRABC}$ shared between Reference ($R_YR$), Bob ($BYC$) and Alice ($A$) such that $\Phi \in  \ball{3\eps_1+5\eps_2 + 2\delta_1+2\delta_2}{\Psi}$.}
\end{theorem}
\begin{proof}

We prove this theorem in three steps. First we recall a proposition that characterizes the number of qubits required when the channel between the sender and receiver is noiseless. Next, we use quantum teleportation to replace the required qubit communication  in the first step into classical communication (along with additional maximally entangled state shared between Alice and Bob). Finally, the classical communication is simulated with an entanglement assisted protocol over the channel $\mathcal{N}_{J\to K}$.

Using quantum teleportation, the number of qubits required in Proposition~\ref{averagecase} can be transmitted to Bob with $2\ell_q$ classical bits. We divide these bits into 
$$\frac{2\ell_q}{C^{\delta_1/ 2\ell_q}(\cN_{J \rightarrow K}) + 2\log\delta_2 - 2\log (2\ell_q)}$$
 blocks, with each block containing 
 $$C^{\delta_1/ 2\ell_q}(\cN_{J \rightarrow K}) + 2\log\delta_2 - 2\log (2\ell_q)$$
  bits.

For any such block $b$, Alice and Bob employ the protocol from \cite[Theorem 1]{AnshuJW17CC} for entanglement-assisted communication over the channel $\cN_{J \rightarrow K}$. As shown in \cite[Theorem 1]{AnshuJW17CC}, the probability that Bob incorrectly decodes any string is upper bounded by $\frac{\delta_1 + \delta_2}{\ell_q}$. Since the number of blocks is at most $2\ell_q$, the overall error is upper bounded by 
$$2\ell_q\cdot\frac{\delta_1 + \delta_2}{\ell_q} = 2\delta_1+2\delta_2.$$

{The number of channel uses is equal to the number of blocks}. Using the error guarantee from Proposition \ref{averagecase} and triangle inequality for the purified distance (Fact \ref{fact:trianglepurified}), the theorem follows. 

\end{proof}

We require the following proposition in Theorem~\ref{worstcasenoiseminimax}.
\begin{proposition}[Achievability bound]
\label{worstcaseminmax}
Fix $\eps_1,\eps_2\in (0,1)$. There exists a $(n' ,3\eps_1+5\eps_2)$-${\downarrow\rm{QCP}}$ code for the quantum state $\Psi^y_{RABC}$ with the number of uses $n'$ of the noiseless qubit channel upper bounded by
\begin{multline*}
\max_{p(y)}\inf_{\sigma_C}\frac{1}{2}\left(\inf_{\Psi'\in \ball{\eps_1}{\Psi}}\dmax{\Psi'_{R_YRAC}}{\Psi'_{R_YRA}\otimes \sigma_C}\right. \\
 - \biggl.\dmineps{\Psi_{AC}}{\Psi_{A}\otimes \sigma_C}{\eps_2}\biggr) + \log\left(\frac{1}{\varepsilon_1\cdot \eps_2}\right),
\end{multline*}
where $\ket{\Psi}_{R_YYRABC}=\sum_{y}\sqrt{p(y)}\ket{y,y}_{R_YY}\otimes\ket{\Psi^y}_{RABC}$.
\end{proposition}
\begin{proof}
Fix a distribution $p(y)$. Consider a unitary protocol $\cP$ for the quantum state redistribution of the quantum state $\sum_{y}\sqrt{p(y)}\ket{y,y}_{R_YY}\otimes\ket{\Psi^y}_{RABC}$ as given in Proposition \ref{averagecase}. It starts with a shared state $\ket{\theta}_{E_AE_B}$, followed by a unitary operation $U$ on all registers other than $RR_Y$. After this, the quantum state is close to $\sum_{y}\sqrt{p(y)}\ket{y,y}_{R_YY}\otimes\ket{\Psi^y}_{RABC}\otimes \ket{\theta'}_{F_AF_B}$. Define the error of the protocol as
\begin{multline*}\err_{p}(\cP) = 1- \biggl |\sum_y p(y) \biggr. \\
\biggl. \bra{y}\bra{\Psi^y}_{RABC}\bra{\theta'}_{F_AF_B} U \ket{y}_{R_YY}\otimes\ket{\Psi^y}_{RABC}\ket{\theta}_{E_AE_B}\biggr |^2,
\end{multline*}
 which is the square of the purified distance from the final state. 
This can be rewritten as 
\begin{equation*}
\err_{p}(\cP) = 1- \left(\sum_y p(y) A_y)^2 - (\sum_y p(y) B_y\right)^2,
\end{equation*}
where $A_y$ ($B_y$) is the real (imaginary) part of $\bra{y}\bra{\Psi^y}_{RABC}\bra{\theta'}_{F_AF_B} U \ket{y}_{R_YY}\otimes\ket{\Psi^y}_{RABC}\ket{\theta}_{E_AE_B}$. This function is concave in $p$. One can take a convex combination of unitary protocols using shared randomness. Let $\{\cP_i\}_i$ be the set of all unitary protocols with quantum communication cost at most 
\begin{multline*}
\max_{p(y)}\inf_{\sigma_C}\frac{1}{2}\left(\inf_{\Psi'\in \ball{\eps_1}{\Psi}}\dmax{\Psi'_{R_YRAC}}{\Psi'_{R_YRA}\otimes \sigma_C} \right. \\
 - \biggl. \dmineps{\Psi_{AC}}{\Psi_{A}\otimes \sigma_C}{\eps_2}\biggr) + \log\left(\frac{1}{\varepsilon_1\cdot \eps_2}\right),
\end{multline*}
and bounded dimension of the state shared between Alice and Bob. It can be verified that the protocol constructed in Proposition \ref{averagecase} has this property. Let $\cP$ be any protocol obtained by using shared randomness to run protocol $\cP_i$ with probability $r_i$. Define 
$$\err_p(\cP) \defeq \sum_i r_i \err_p(\cP_i). $$ Thus, the function $\err_p(\cP)$ is linear (and hence convex) in $\cP$ and concave in $p$. The set of protocols $\cP$ are convex and compact as all the unitary protocols $\cP_i$ act on registers of dimension at most $D$, where $D$ is an integer that is a function of the input states $\{\ket{\Psi^y}_{RABC}\}_y$. Furthermore, the set of probability distributions $p$ is also convex and compact. Thus, we can apply the minimax Theorem \ref{fact:minimax} to conclude that 
$$\max_{p}\min_{\cP}\err_p(\cP) = \min_{\cP}\max_{p}\err_p(\cP) \leq (3\eps_1+5\eps_2)^2.$$ Thus, there exists a protocol $\cP$ that makes an error of at most $3\eps_1 + 5\eps_2$ in purified distance for every distribution $p$. In particular, we can choose $p$ to be point distributions, leading to the desired worst case bound. This completes the proof. 
\end{proof}

\begin{theorem}[Achievability bound]
\label{worstcasenoiseminimax}
Fix $\eps_1,\eps_2, \delta_1, \delta_2 >0$ and a quantum channel $\cN_{J \rightarrow K}$. Define $$C^{\delta_1}(\cN_{J \rightarrow K}) \defeq \max_{\ketbra{\theta}_{JJ'}}\dmineps{\cN_{J\rightarrow K}(\theta_{JJ'})}{\cN_{J\rightarrow K}(\theta_{J})\otimes \theta_{J'}}{\delta_1}.$$  There exists a $(n,3\eps_1+5\eps_2 + 2\sqrt{2\delta_1 + 2\delta_2})$-${\downarrow\rm{QCP}}$ code for the quantum state $\Psi^y_{RABC}$ with $n$, that is the number of uses of the quantum channel $\cN_{J \rightarrow K}$ between Alice and Bob, upper bounded by 

\begin{eqnarray*}
\frac{\bar{\ell}_q} {C^{\delta_1/2\bar{\ell}_q}(\cN_{J \rightarrow K}) + 2\log\delta_2 - 2\log(2\bar{\ell}_q)},
\end{eqnarray*}
where 
\begin{multline*}\bar{\ell}_q \defeq \frac{1}{2}\inf_{\sigma_C}\left(\max_y\inf_{\Psi'\in \ball{\eps_1}{\Psi^y}}\dmax{\Psi'_{RAC}}{\Psi'_{RA}\otimes \sigma_C} \right.\\
- \biggl.\dmineps{\Psi_{AC}}{\Psi_{A}\otimes \sigma_C}{\eps_2}\biggr)+ \log\left(\frac{1}{\varepsilon_1\cdot \eps_2}\right) .
\end{multline*}

\suppress{In other words, there exists an entanglement-assisted one-way protocol $\cP$, which takes as input $\Psi^y_{RABC}$ shared between three parties Reference ($R$), Bob ($B$) and Alice ($AC$) and outputs a state $\Phi^y_{RABC}$ shared between Reference ($R$), Bob ($BC$) and Alice ($A$) such that $\Phi^y \in  \ball{3\eps_1+5\eps_2 + 2\delta_1 + 2\delta_2}{\Psi^y}$.}
\end{theorem}

\begin{proof}

Our proof idea follows by adding the use of the minimax theorem (Fact \ref{fact:minimax}) to the proof of Theorem~\ref{averagecasenoise}. We start with proving a proposition that characterizes the number of qubits required for transmitting $C$ of the state $\Psi^y_{RABC}$ when the channel between the sender and receiver is error free. The minimax theorem then allows us to relate this case to the worst case error.  The next two steps are to use quantum teleportation to convert the required qubit communication into classical communication, and to simulate the classical communication with a noisy entanglement-assisted protocol over the channel $\mathcal{N}_{J\to K}$.

%We invoke Proposition~\ref{worstcaseminmax} to obtain the result
The result can be obtained by using quantum teleportation to convert the qubit communication in Proposition~\ref{worstcaseminmax} into classical communication, followed by simulating the classical communication with a noisy entanglement-assisted protocol over the channel $\mathcal{N}_{J\to K}$ (as in the proof of Theorem~\ref{averagecasenoise}).

\end{proof}

We remark that Theorem~\ref{worstcasenoiseminimax} only gives us non-explicit protocols. We can however also give explicit protocols for the worse case error, but with a slightly loose upper bound. The protocol for the noiseless case, as constructed below uses the bounds given in \cite{AnshuJW17SR} and the construction of the union of projectors given in \cite{AnshuJW17comp}. 

\begin{proposition}[Achievability bound for ideal qubit channel]
\label{worstcase}
Fix $\eps_1,\eps_2\in (0,1)$.  There exists a $(n',3\eps_1+5\eps_2)$-${\downarrow\rm{QCP}}$ code for the quantum state $\Psi^y_{RABC}$ with the number of $n'$ uses  of the ideal qubit channel between Alice and Bob, upper bounded by 
\begin{multline*}
\inf_{\sigma_C}\frac{1}{2}\left( \max_y\inf_{\Psi'\in \ball{\eps_1}{\Psi^y}}\dmax{\Psi'_{RAC}}{\Psi'_{RA}\otimes \sigma_C} \right. \\
 - \biggl.\min_y\dmineps{\Psi^y_{AC}}{\Psi^y_{A}\otimes \sigma_C}{\eps_2}\biggr) \\ + \log 2|\cY|\cdot \log\log2|\cY| + \log\left(\frac{1}{\varepsilon_1\cdot \eps_2}\right).
\end{multline*}
If the register $A$ is trivial, then there exists a $(n',3\eps_1)$-${\downarrow\rm{QCP}}$ code for the quantum state $\Psi^y_{RABC}$ with the number of $n'$ uses  of the ideal qubit channel between Alice and Bob, upper bounded by
\begin{eqnarray*}
\frac{1}{2}\left( \inf_{\sigma_C}\max_y\inf_{\Psi'\in \ball{\eps_1}{\Psi^y}}\dmax{\Psi'_{RC}}{\Psi'_{R}\otimes \sigma_C}\right) + \log\left(\frac{1}{\varepsilon_1}\right),
\end{eqnarray*}
which is independent of $|\cY|$.
\end{proposition}

\begin{proof}
Let $\Pi^y_{AC}$ be the operator such that 
$$\Tr(\Pi_{AC}^y(\Psi^y_A\otimes \Psi^y_C)) \leq 2^{- \dmineps{\Psi^y_{AC}}{\Psi^y_{A}\otimes \sigma_C}{\eps_2}},$$
 and 
 $$\Tr(\Pi_{AC}^y\Psi^y_{AC}) \geq 1-\eps_2.$$ Using Naimark's theorem to extend $\Pi^y_{AC}$ into a projector and then invoking \cite[Theorem 2]{AnshuJW17comp}, there exists an operator $\Pi^*_{AC}$ such that $\Tr(\Pi^*_{AC}\Psi^y_{AC})\geq 1-2\eps_2$ and 
\begin{multline*}
\Tr(\Pi^*_{AC}(\Psi^y_A\otimes \Psi^y_C)) \\  \leq 2^{- \min_y\dmineps{\Psi^y_{AC}}{\Psi^y_{A}\otimes \sigma_C}{\eps_2} + 2\log 2|\cY|\cdot \log\log2|\cY|},
\end{multline*}
 for all $y$. 

We will now construct a reversible protocol $\cP_2$ for a reversed task where Alice, Bob and Reference start with the state $\Psi^y_{RABC}$ shared between Reference ($R$), Bob ($BC$) and Alice ($A$) and end with a state $\Phi'^y_{RABC}$ shared between Reference ($R$), Bob ($B$) and Alice ($AC$) such that $\Phi'^y \in  \ball{3\eps_1+5\eps_2}{\Psi^y}$. Further, Bob knows $y$ and Alice is unaware of it. It can be verified that reversing this protocol leads to the desired protocol $\cP$. 

The construction of the protocol $\cP_2$ directly follows from the construction given in \cite[Theorem 1]{AnshuJW17SR} and the operator $\Pi^*_{AC}$ constructed above. {In more details, the protocol is composed of a convex-split step and a quantum hypothesis testing step (also termed as the position-based decoding). In the convex-split step, Bob (who is the sender in the protocol $\cP_2$ and is aware of $y$) applies a measurement conditioned on the input $y$ which has 
$$\frac{2^{\max_y\inf_{\Psi'\in \ball{\eps_1}{\Psi^y}}\dmax{\Psi'_{RAC}}{\Psi'_{RA}\otimes \sigma_C}}}{\eps_1}$$
 outcomes. Each outcome determines a register $C$ on Alice's side with which the registers $RA$ are properly correlated. If Alice knew the outcome, she would be able to output the correct quantum state. Instead of sending the outcome, Bob communicates partial information about the outcome. This is done in a manner that Alice is uncertain about only 
 $$\eps_2\cdot 2^{\min_y\dmineps{\Psi^y_{AC}}{\Psi^y_{A}\otimes \sigma_C}{\eps_2}- 2\log 2|\cY|\cdot \log\log2|\cY|}$$ outcomes of Bob. To estimate the correct outcome with small error, Alice performs the quantum hypothesis testing measurement using the operator $\Pi^*_{AC}$. Overall, Bob manages to save in communication, with a small increase in error at Alice's end.} The analysis of the protocol and the proof of correctness follow \cite[Theorem 1]{AnshuJW17SR}.

If the register $A$ is trivial, then the desired bound is obtained through a protocol $\cP_3$ where Alice does not perform any quantum hypothesis testing. The analysis of the protocol follows from the protocol for quantum state splitting given in \cite{ADJ14}.
\end{proof}

The noisy version of Proposition \ref{worstcase} is now as follows and its proof is similar to the proof of Theorem \ref{averagecasenoise}.

\begin{theorem}[Achievability bound]
\label{worstcasenoise}
Fix $\eps_1,\eps_2, \delta_1, \delta_2 >0$ and a quantum channel $\cN_{J \rightarrow K}$. Define $$C^{\delta_1}(\cN_{J \rightarrow K}) \defeq \max_{\ketbra{\theta}_{JJ'}}\dmineps{\cN_{J\rightarrow K}(\theta_{JJ'})}{\cN_{J\rightarrow K}(\theta_{J})\otimes \theta_{J'}}{\delta_1}.$$ 
 There exists a $(n,3\eps_1+5\eps_2 + 2\sqrt{2\delta_1 + 2\delta_2})$-${\downarrow\rm{QCP}}$ code for the quantum state $\Psi^y_{RABC}$ with $n$, that is the number of uses of the quantum channel $\cN_{J \rightarrow K}$ between Alice and Bob, upper bounded by 
$$\max\left(\frac{2\ell^*_q}{C^{\delta_1/ 2\ell^*_q}(\cN_{J \rightarrow K}) + 2\log\delta_2 - 2\log(2\ell^*_q)}, 1\right),$$ where 
\begin{multline*}
\ell^*_q =  \frac{1}{2}\inf_{\sigma_C}\left( \max_y\inf_{\Psi'\in \ball{\eps_1}{\Psi^y}}\dmax{\Psi'_{RAC}}{\Psi'_{RA}\otimes \sigma_C} \right. \\
 - \biggl. \min_y\dmineps{\Psi^y_{AC}}{\Psi^y_{A}\otimes \sigma_C}{\eps_2}\biggr) \\ + \log 2|\cY|\cdot \log\log2|\cY| + \log\left(\frac{1}{\varepsilon_1\cdot \eps_2}\right).
\end{multline*}
If the register $A$ is trivial, then $\Phi^y \in  \ball{3\eps_1+ 2\sqrt{2\delta_1 + 2\delta_2}}{\Psi^y}$ and $\ell^*_q$ can be chosen to be equal to  
\begin{eqnarray*}
\inf_{\sigma_C}\max_y\inf_{\Psi'\in \ball{\eps_1}{\Psi^y}}\dmax{\Psi'_{RC}}{\Psi'_{R}\otimes \sigma_C} + 2\log\left(\frac{1}{\varepsilon_1}\right),
\end{eqnarray*}
which is independent of $|\cY|$.
\end{theorem}

\begin{remark}
\label{remark:chuse}
The error parameter $\frac{\delta_1}{2\ell^*_q}$ as appearing in $C^{\delta_1/ 2\ell^*_q}(\cN_{J \rightarrow K})$ can be improved to $\delta_1$ if it is known that the number of channel uses is one. In such a scenario, the expression 
$$\frac{2\ell^*_q}{C^{\delta_1/ 2\ell^*_q}(\cN_{J \rightarrow K}) + 2\log\delta_2 - 2\log(2\ell^*_q)}$$ is improved to 
$$\frac{2\ell^*_q}{C^{\delta_1}(\cN_{J \rightarrow K}) + 2\log\delta_2}.$$ 
\end{remark}

\subsection{Asymptotic and i.i.d analysis}

In the asymptotic and i.i.d. setting, we have the following result.
\begin{proposition}[Asymptotic and i.i.d. bound]
\label{averageasymptoticiid}
Fix $\eps, \delta \in (0,1)$ and a quantum channel $\cN_{J \rightarrow K}$. Define $$C(\cN_{J \rightarrow K}) \defeq \max_{\ketbra{\theta}_{JJ'}}\mutinf{J'}{K}_{\cN_{J\rightarrow K}(\theta_{JJ'})}.$$
There exists a $N$ large enough and a $(n, \eps)$-$\overline{\rm{QCP}}$ code for the quantum state $\ket{\Psi}^{\otimes N}_{R_YYRABC}$ with $n$ (the number of uses of the quantum channel $\cN_{J \rightarrow K}$ between Alice to Bob) upper bounded by 
\begin{eqnarray*}
N\frac{\condmutinf{R_YR}{C}{BY}_{\Psi} + \delta}{C(\cN_{J \rightarrow K}) -\delta}.
\end{eqnarray*}
Furthermore, for every $\eps, \delta \in (0,1)$, there exists a $N$ large enough such that for any $(n, \eps)$-$\overline{\rm{QCP}}$ code for the quantum state $\ket{\Psi}^{\otimes}_{R_YYRABC}$, the number of uses $n$ of the channel is at least
\begin{eqnarray*}
N\frac{\condmutinf{R_YR}{C}{BY}_{\Psi} - \delta}{C(\cN_{J \rightarrow K}) + \delta}.
\end{eqnarray*}
\end{proposition}
\begin{proof}
{The achievability result is a direct consequence of the one-shot result in Theorem \ref{averagecasenoise} and its asymptotic i.i.d. analysis that can be performed using Fact \ref{dmaxequi}. An alternate approach (which is similar to Theorem \ref{averagecasenoise}) is to employ the protocols in \cite{Devatakyard, YardD09, BennettSST02}, as follows.  We consider the quantum state redistribution protocol for $\Psi_{R_YRCBY}$ \cite{Devatakyard, YardD09}, which requires a classical communication at a rate of 
$$\condmutinf{R_YR}{C}{BY}_{\Psi}+\delta$$
 for large enough $N$. In order to communicate the classical bits, the channel $\cN_{J\rightarrow K}$ must be used. But, the classical communication through this channel is possible at a rate of $C(\cN_{J \rightarrow K}) -\delta$ bits \cite{BennettSST02}. From this, the claimed number of channel uses follows.} 

For the converse, we use the quantum Reverse Shannon Theorem \cite{BDHSW14, Renner11}, which says that for every $\eps, \delta \in (0,1)$, there exists a $k$ large enough such that the $k$ uses of a channel $\cN_{J \rightarrow K}$ can be simulated with communication cost 
$$k(C(\cN_{J \rightarrow K}) + \delta).$$ Further, from the converse given in \cite{Devatakyard}, there exists a $N$ large enough such that any protocol achieving quantum state redistribution of $\Psi^{\otimes N}_{R_YYRABC}$ requires at least 
$$N(\condmutinf{R_YR}{C}{BY}_{\Psi} - \delta)$$ bits of communication. If the number of uses of the channel is smaller than 
$$N\frac{\condmutinf{R_YR}{C}{BY}_{\Psi} - \delta}{C(\cN_{J \rightarrow K}) + \delta},$$ then we reach a contradiction. This proves the result.  
\end{proof}

\section{Applications}
\label{sec:ptop}

\subsection{One-shot Alpha-bit capacity with entanglement assistance}

There are two parties Alice  and Bob. Fix a Hilbert space $\cH_Q$ on register $Q$ and a subspace $S\subset \cH_Q$ such that $|S| = |Q|^{\alpha}$. Alice and Reference share a quantum state $\ket{\Psi(S)}_{RQ}$, where $\Psi(S)_Q$ is maximally mixed in the subspace $S$. Alice wants to communicate the register $Q$ to Bob. Further, Alice is unaware of $S$, except for the value of $\alpha$. To accomplish this task Alice  and Bob  also share entanglement between them.  We now make the following definition:
\begin{definition}
\label{qcode}
Let $\ket{\theta}_{E_AE_B}$ be the state shared between Alice and Bob. A $(\log|Q|, \eps, \alpha )$-entanglement assisted code for quantum communication over the quantum channel $\cN_{ A \to B}$ consists of 
\begin{itemize}
\item An encoding operation $\cE: QE_A \rightarrow A$ for Alice that does not depend on $S$.  
\item A decoding operation $\cD : BE_B\rightarrow Q'$ for Bob, such that $Q'\equiv Q$ and
\begin{equation}
\Pur\left(\ketbra{\Psi(S)}_{RQ'}, \omega_{RQ'} \right) \leq \eps,
\end{equation}
where $\omega_{RQ'}= \cD\circ\cN_{A\to B}\circ\cE\left(\ketbra{\Psi(S)}_{RQ}\right)$.
\end{itemize}
\end{definition}

Now we give a one-shot achievability protocol for the task defined in Definition \ref{qcode}. It follows from a simple application of Theorem \ref{worstcasenoise} and Remark \ref{remark:chuse}.
\begin{theorem}
\label{theo:qachieve}
Let $\cN_{A \to B }$ be the quantum channel and let $\eps,\delta \in (0,1)$. Let $A'\equiv A$ be a purifying register. Then, for any $|Q|$ satisfying 
\begin{multline*}
\log|Q| \leq \\  \frac{1}{1+\alpha}\max _{\ketbra{\psi}_{A A'}} \dmineps{\cN_{A \to B } (\ketbra{\psi}_{AA'})}{\cN_{A\to B}(\psi_{A}) \otimes \psi_{A'}}{\eps}\\ - \frac{4}{1+\alpha}\log \frac{1}{\delta},
\end{multline*}
there exists a $(\log|Q|, 2\eps + 2\sqrt{5\delta}, \alpha)$ entanglement assisted code for quantum communication over the quantum channel $\cN_{A \to B }.$
\end{theorem}
\begin{proof}
We invoke Theorem \ref{worstcasenoise} with register $A$ trivial,  $\cY$ as the set of all subspaces $S$ of $\cH_Q$\footnote{While this set is uncountable, we can also consider its finite version by choosing appropriate covering nets and allowing a small error (going to zero).} of dimension $|Q|^{\alpha}$, $\Psi^y_{RABC}$ as the collection of quantum states $\ket{\Psi(S)}_{RQ}$ and $\eps_1, \delta_2 = \delta, \delta_1=\eps$.  We have that 
\begin{eqnarray*}
&& \inf_{\sigma_Q}\max_S\inf_{\Psi'\in \ball{\delta}{\Psi(S)}}\dmax{\Psi'_{RQ}}{\Psi'_{R}\otimes \sigma_Q}\\
&& \leq \max_S\dmax{\Psi(S)_{RQ}}{\Psi(S)_{R}\otimes \frac{\id}{|Q|}} \\
&& = (1+\alpha)\log|Q|.
\end{eqnarray*}
Since the number of channel uses is one, following Remark \ref{remark:chuse}, the maximum possible value of $|Q|$ is obtain by setting 
\begin{multline*}
\frac{\inf_{\sigma_Q}\max_S\inf_{\Psi'\in \ball{\delta}{\Psi(S)}}\dmax{\Psi'_{RQ}}{\Psi'_{R}\otimes \sigma_Q} + 2\log\left(\frac{1}{\delta}\right)}{C^{\eps}(\cN_{A \rightarrow B}) + 2\log\delta}\\ \leq 1.
\end{multline*} 
This is satisfied if 
$$\frac{ (1+\alpha)\log|Q|+ 2\log\left(\frac{1}{\delta}\right)}{C^{\eps}(\cN_{A \rightarrow B}) + 2\log\delta}\leq 1,$$ which completes the proof. 
\end{proof}

An immediate corollary of this is to recover the entanglement assisted Alpha-bit capacity of \cite{1706.09434}. 
\begin{corollary}
\label{cor:qachieve}
Let $\cN_{A \to B }$ be the quantum channel and $\eps,\delta \in (0,1)$. There exists a $n$ large enough such that for any $q'$ satisfying
$$q' \leq  \frac{1}{1+\alpha}\max _{\ketbra{\psi}_{A A'}}\mutinf{B}{A'}_{\cN_{A \to B } (\ketbra{\psi}_{AA'})},$$
there exists a $(n(q' - \delta), \eps, \alpha)$- entanglement assisted code for quantum communication over the quantum channel $\cN^{\otimes n}_{A \to B }.$ 
\end{corollary} 
\begin{proof}
Let $n>0$ be an integer to be chosen later. Applying Theorem \ref{theo:qachieve} to the quantum channel $\cN^{\otimes n}_{A\to B}$ , 
there exists a $(\log|Q|, 2\eps + 2\sqrt{5\delta}, \alpha)$ entanglement assisted code for quantum communication over the quantum channel $\cN_{A \to B }.$  
\begin{multline*}
\log|Q| \leq  \frac{1}{1+\alpha}\max _{\ketbra{\psi}_{A^nA'^n}} \\\dmineps{\cN^{\otimes n}_{A \to B } (\ketbra{\psi}_{A^nA'^n})}{\cN^{\otimes n}_{A\to B}(\psi_{A^n}) \otimes \psi_{A'^n}}{\eps} \\
- \frac{4}{1+\alpha}\log \frac{1}{\delta}.
\end{multline*}
Restricting the maximization to product states $\psi_{AA'}^{\otimes n}$ and applying Facts \ref{dmaxequi}, \ref{gaussianupper}, we conclude that it suffices to have
\begin{multline*}
\log|Q| \leq  \frac{1}{1+\alpha}\Biggl(n\cdot\max _{\ketbra{\psi}}\mutinf{B}{A'}_{\cN_{A \to B } (\ketbra{\psi}_{AA'})} \Biggr. \\
- \Biggl. O\left(\sqrt{n \log\frac{1}{\eps}}\right)\Biggr) \\
= \frac{n}{1+\alpha}\Biggl(\max _{\ketbra{\psi}}\mutinf{B}{A'}_{\cN_{A \to B } (\ketbra{\psi}_{AA'})} \Biggr. \\
 -\Biggl. O\left(\sqrt{\frac{\log\frac{1}{\eps}}{n}}\right)\Biggr).
\end{multline*}
Let $n$ be chosen large enough such that $\delta \geq O\left(\sqrt{\frac{\log\frac{1}{\eps}}{n}}\right)$. This completes the proof.
\end{proof}

\subsection{Alpha-bit capacity without entanglement assistance}

We will start with an average case version of the Alpha-bit transmission, to provide a simple introduction to the protocol. The worst case version will build upon this protocol. 

\begin{definition}[Uniform average case of Alpha-bit]
\label{avunqcode}
Fix a register $Q$. A $(\log|Q|, \eps, \alpha, n)$-entanglement unassisted average case code for quantum communication over the quantum channel $\cN_{J \to K}$ consists of $n$ registers $Q_i\equiv Q$, subspaces $S_1, S_2, \ldots S_n$ of dimension $|Q|^\alpha$ each, and quantum states $\ket{\Psi(S_i)}_{RQ}$ maximally entangled in the subspace $S_i$ of $Q_i$ with a suitable reference system $R$, such that there exists an integer $m$
\begin{itemize}
\item An encoding operation $\cE: Q^n \rightarrow J^m$ for Alice that does not depend on $S_1, \ldots S_n$.  
\item A decoding operation $\cD : J^m\rightarrow Q'^n$ for Bob, such that $Q'\equiv Q$ and
$$\mathbb{E}_{S_1, S_2, \ldots S_n}\Pur\left(\otimes_i\ketbra{\Psi(S_i)}_{RQ'_i}, \omega_{R^nQ'^n} \right)\leq \eps, $$
where $\omega_{R^nQ'^n}=\cD\circ\cN^{\otimes m}\circ\cE\left(\otimes_i\ketbra{\Psi(S_i)}_{RQ_i}\right)$
and the average is taken according to the uniform distribution.
\end{itemize}

\end{definition}

We will also use the protocol for entanglement assisted quantum communication as given in \cite{DevetakHW04}.

\begin{theorem}[Entanglement assisted quantum capacity, \cite{DevetakHW04, BennettSST02}]
\label{qcapacitytheo}
Fix a quantum channel $\cN_{J\to K}$ and the complementary channel $\cN^c_{J \to L}$. Let $\psi_{JJ'}$ be an arbitrary quantum state and $(W, E)$ be any pair satisfying 
\begin{eqnarray*}
W &\leq& \frac{1}{2}I(J':K)_{\cN_{J\to K}(\psi_{JJ'})}, \\
V &\geq& \frac{1}{2}I(J':L)_{\cN^c_{J\to L}(\psi_{JJ'})}.
\end{eqnarray*}
There exists a real $E_2 \geq 0$ such that the following holds. For every $\eps, \delta>0$, there exists a $n$ large enough such that there exists a one-way protocol for communicating $n(W-\delta)$ qubits with error $\eps$, number of ebits of maximally entangled state shared in the beginning equal to $n(V+ E_2+\delta)$ and the number of ebits of maximally entangled state returned equal to $n(E_2 - \delta)$. 
\end{theorem}

We use above results to prove the following theorem.
\begin{theorem}
\label{unassistedav}
Fix a quantum channel $\cN_{J\to K}$ and an $\alpha\in (0,1)$.  Let $\psi_{JJ'}$ be an arbitrary quantum state and define 
\begin{eqnarray*}
W &=& \frac{1}{2}I(J':K)_{\cN_{J\to K}(\psi_{JJ'})}, \\
V &=& \frac{1}{2}I(J':L)_{\cN^c_{J\to L}(\psi_{JJ'})}, \\
Y &=& \max\left(- \condent{J'}{K}_{\cN_{J\to K}(\psi_{JJ'})}, 0\right).
\end{eqnarray*} For every $\eps > 0$, there exist $n, d$ large enough such that there exists a $(\log d, 4\eps, \alpha, n)$-entanglement unassisted average case code such that $n\log d$ divided by the number of channel use (or number of $\alpha$-bits transmitted per channel use) is equal to 
\vspace{1mm}
\begin{itemize}
\item $\frac{2}{1+\alpha}W,$ if $\frac{W}{V}\geq \frac{1+\alpha}{1-\alpha}$ (for this case, $\frac{2}{1+\alpha}W \leq \frac{Y}{\alpha}$) or \vspace{2mm}
\item $\frac{Y}{\alpha},$ if $\frac{W}{V}< \frac{1+\alpha}{1-\alpha}$ (for this case, $\frac{2}{1+\alpha}W \geq \frac{Y}{\alpha}$). 
\end{itemize}
\end{theorem}
\begin{proof}

Let $k$ be an arbitrary positive integer. Consider the pure quantum state $$\ket{\Psi}_{RQR_SS}\defeq \sum_S\sqrt{\mu(S)} \ket{\Psi(S)}_{RQ}\ket{S, S}_{R_S, S}.$$
As shown in \cite{Berta14}, there is an entanglement assisted protocol for quantum state merging of $\ketbra{\Psi}_{RQR_SS}^{\otimes m}$ with error $\frac{\eps}{k}$, the quantum communication cost 
$$m\left(\frac{1+\alpha}{2}\log d +  4\log \frac{mk}{\eps}\right),$$ initial entanglement of $0$ ebits and final entanglement of $$m\left(\frac{1-\alpha}{2}\log d - 4\log \frac{mk}{\eps}\right)$$ ebits.

From Theorem \ref{qcapacitytheo}, there exists a $m'$ large enough and a protocol for communicating $m'(W- \delta)$ qubits with initial entanglement of $m'(E_2 + V + \delta)$ ebits and final entanglement of $m'(E_2 - \delta)$ ebits with $m$ uses of the channel $\cN$ and error $\frac{\eps}{k}$.

There exists a $m''$ large enough such that there exists a protocol for communicating $m''(Y - \delta)$ qubits over the channel $\cN$ with error $\frac{\eps}{k}$. Let $m',m''$ be such that the following inequalities are satisfied. 
\begin{equation}
\label{unassistedconstraintav}
m\left(\frac{1+\alpha}{2}\log d +  4\log \frac{mk}{\eps}\right) - m'(W- \delta) \leq m''(Y - \delta),
\end{equation}  
and 
\begin{equation}
\label{ebitincreaseav}
m\left(\frac{1-\alpha}{2}\log d -  4\log \frac{mk}{\eps}\right) \geq m'(V + 2\delta).
\end{equation}
 
The choice of $m', m''$ is made later. The protocol is as follows.
\begin{itemize}
\item The protocol starts with communicating $m'(E_2 + V + \delta)$ ebits through $O(m + m')$ uses of the channel $\cN$ and error $\frac{\eps}{100}$. \vspace{2mm}
\item If $\frac{W}{V}\leq \frac{1+\alpha}{1-\alpha}$,
\begin{itemize} 
\item Alice aims to communicate 
$$m\left(\frac{1+\alpha}{2}\log d +  4\log \frac{mk}{\eps}\right)$$ qubits to Bob using the quantum state merging protocol. The first $m'(W- \delta)$ qubits are communicated using the entanglement-assisted protocol, using the shared entangled state of $m'(E_2 + V + \delta)$ ebits and the remaining 
$$m\left((\beta+\alpha)\log d +  4\log \frac{mk}{\eps}\right) - m'(W- \delta)$$ qubits are communicated using the entanglement unassisted protocol for quantum communication at the rate of $Y$ qubits per channel use. 
\end{itemize} \vspace{1mm}
\item If $\frac{W}{V}\geq \frac{1+\alpha}{1-\alpha}$,
\begin{itemize} \vspace{1mm}
\item Alice aims to communicate 
$$m\left(\frac{1+\alpha}{2}\log d +  4\log \frac{mk}{\eps}\right)$$ qubits to Bob using the quantum state redistribution protocol. She communicates them using $m'$ uses of the channel with the entanglement unassisted protocol. 
\end{itemize}
\item Alice and Bob repeat this protocol till $k$ rounds. It is ensured that the number of ebits for the next round of the protocol is enough by Equation \ref{ebitincreaseav}.
\end{itemize}

\noindent {\bf Error Analysis:} The overall error is at most 
$$\frac{\eps}{100} + k(\frac{3\eps}{k}) \leq 4\eps.$$ 

\vspace{0.1in}

\noindent {\bf Rate of communication:} The number of channel uses for transmitting $km$ copies of $\alpha$-dits is 
$$km'+km'' + O(m'+m'') = \left(1+ \frac{O(1)}{k}\right)\cdot (km'+km'').$$   Thus, the number of $\alpha$-dits transmitted per channel use is 
$$ \left(1+ \frac{O(1)}{k}\right)\frac{m}{m'+m''}.$$ This implies that the number of $\alpha$-bits transmitted per channel use is 
$$ \left(1+ \frac{O(1)}{k}\right)\frac{m\log d}{m'+m''}.$$ The achievable rate in the asymptotic and i.i.d. setting is then obtained by considering 
\begin{eqnarray*}
&&\lim_{\delta \rightarrow 0}\lim_{k\rightarrow \infty}\lim_{m,m' \rightarrow \infty}\lim_{d\rightarrow \infty} \left(1+ \frac{O(1)}{k}\right)\frac{m\log d}{m'+m''} \\ 
&=& \lim_{\delta \rightarrow 0}\lim_{m,m' \rightarrow \infty}\lim_{d\rightarrow \infty} \frac{m\log d}{m'+m''}.
\end{eqnarray*}
 To upper bound this quantity, we consider the following two cases, setting $\delta=0$ and $\log\frac{mk}{\eps}$ to $0$ below (as the latter term is subsumed by $\log d$ terms). 
\vspace{1mm}
\begin{itemize}
\item If it holds that $\frac{1-\alpha}{V} \geq \frac{1+\alpha}{W}$, then we set $m''=0$. We set 
$$m'W = m(\frac{1+\alpha}{2}\log d) \geq m(\frac{1+\alpha}{2}\log d).$$ This satisfies both Equations \ref{unassistedconstraintav} and \ref{ebitincreaseav}. The number of $\alpha$-bits transmitted per channel use is equal to $\frac{2}{1+\alpha}W$. 
\vspace{1mm}
\item If it holds that $\frac{1-\alpha}{V} < \frac{1+\alpha}{W}$ and $W>V$, we saturate both Equations \ref{unassistedconstraintav} and \ref{ebitincreaseav} to obtain
\begin{eqnarray*}
m\left(\frac{1+\alpha}{2}\log d\right) - m'W &=&  m''Y, \\
 - m\left(\frac{1-\alpha}{2}\log d \right) + m'V &=& 0.
\end{eqnarray*}
Adding the equations, we obtain
$$m\alpha\log d - m'(W-V) = m'' Y.$$
Using the relation $W-V = Y$ for $Y>0$, this gives us 
$$m\alpha\log d = (m'+m'')Y \implies \frac{m\log d}{m'+m''} = \frac{Y}{\alpha}.$$
\item If $W< V$, then $Y=0$, by definition. No value of $m',m''$ satisfies Equations \ref{unassistedconstraintav} and \ref{ebitincreaseav}, unless if $m=0$. 
\end{itemize}

This completes the proof.
\end{proof}

Now we are in a position to define the worst case version.

\begin{definition}[Entanglement unassisted transmission of Alpha-bit]
\label{unqcode}
Fix a register $Q$. A $(\log|Q|, \eps, \alpha, n)$-entanglement unassisted code for quantum communication over the quantum channel $\cN_{J \to K}$ consists of $n$ registers $Q_i\equiv Q$, subspaces $S_1, S_2, \ldots S_n$ of dimension $|Q|^\alpha$ each, and quantum states $\ket{\Psi(S_i)}_{RQ}$ maximally entangled in the subspace $S_i$ of $Q_i$ with a suitable reference system $R$, such that there exists an integer $m$
\begin{itemize}
\item An encoding operation $\cE: Q^n \rightarrow J^m$ for Alice that does not depend on $S_1, \ldots S_n$.  
\item A decoding operation $\cD : J^m\rightarrow Q'^n$ for Bob, such that $Q'\equiv Q$ and
$$\Pur\left(\otimes_i\ketbra{\Psi(S_i)}_{RQ'_i},\omega_{R^nQ'^n} \right)\leq \eps,$$
where $\omega_{R^nQ'^n}=\cD\cN^{\otimes m}\cE\left(\otimes_i\ketbra{\Psi(S_i)}_{RQ_i}\right)$.
\end{itemize}
\end{definition}

By a careful application of the minimax theorem (Fact \ref{fact:minimax}), we are able to extend Theorem \ref{unassistedav} such that it works for all subspaces. We start with the following proposition.

\begin{proposition}
\label{stateredisttheo}
Fix $\eps\in (0,1)$. There exists a randomness and entanglement-assisted one-way protocol $\cP$, which takes as input $\Psi(S)_{RQ}$ (for any arbitrary subspace $S$) shared between Reference ($R$) and Alice ($Q$) and outputs a state $\Phi(S)_{RQ}$ shared between Reference ($R$) and Bob ($Q$) such that $\Phi(S) \in  \ball{10\eps}{\Psi(S)}$. No ebits of shared state required in the protocol. The number of qubits communicated by Alice to Bob is 
$$\frac{1+\alpha}{2}\log d + 4\log\frac{1}{\eps},$$ and the number of ebits gained by the protocol is $$\frac{1-\alpha}{2}\log d- 4\log\frac{1}{\eps}.$$
\end{proposition}
\begin{proof}
Let $\delta, \delta'\in (0,1)$. We use the protocol from \cite{Berta14}, along with the minimax argument as given in Proposition \ref{worstcaseminmax}. Choose a net $\net$ over the set of subspaces of dimension $d^{\alpha}$, such that the error of approximation between a subspace $T$ and some subspace $S\in \net$ is 
$$\Pur\left(\frac{S}{d^{\alpha}},\frac{T}{d^{\alpha}}\right) \leq \delta.$$ For a given distribution $p(S)$ over the subspaces $S$ chosen from this net, let  $\ket{\Psi}_{R_SSRQ}=\sum_{S}\sqrt{p(S)}\ket{S,S}_{R_SS}\otimes\ket{\Psi(S)}_{RQ}$. As shown in \cite{Berta14}, there exists a entanglement assisted one-way protocol for the quantum state merging of $\ket{\Psi}_{R_SSRQ}$, where the number of qubits communicated by Alice to Bob is  
\begin{eqnarray*}
\frac{1}{2}\left(\hmax{Q}{S}_{\Psi} + \mathrm{H}_{\max}(Q)_{\Psi}\right) + 4\log\left(\frac{1}{\eps}\right),
\end{eqnarray*} 
The number of ebits of shared state required in the protocol is 
$$\left\{\frac{1}{2}\log|Q| - \frac{1}{2}\mathrm{H}_{\max}(Q)_{\Psi}\right\}.$$ The number of ebits consumed in the protocol is
$$\left\{\frac{1}{2}\hmax{Q}{S}_{\Psi} - \frac{1}{2}\mathrm{H}_{\max}(Q)_{\Psi}\right\} + 4\log\frac{1}{\eps}.$$
It can be evaluated that $\hmax{Q}{S}_{\Psi} = \alpha\log d$ and $\mathrm{H}_{\max}(Q)_{\Psi}=\log |Q|$. In particular, if $p(S)$ is a distribution that has full support over the set of subspaces from $\net$, then the support of $\Psi_Q$ is the whole of $\cH_Q$. Thus, $\log|Q| = \log d$. It follows that the number of qubits communicated by Alice to Bob is $(\frac{1+\alpha}{2})\log d$ and the number of ebits gained by the protocol is $-\frac{1-\alpha}{2}\log d$. The error of the protocol is $10\eps$ in the purified distance. 

Now, we apply the minimax theorem (Fact \ref{fact:minimax}, similar to the argument as given in Proposition \ref{worstcaseminmax}) over distributions $p$ for which the probability over any subspace in $\net$ is at least $\delta'$. These distributions form a convex and compact set, and have full support over the set $\net$. This gives a randomness assisted protocol which makes an error of $10\eps + |\net|\delta'$ for every subspace in $\net$. Since 
\begin{eqnarray*}
\Pur(\ketbra{\Psi(S)}_{RQ}, \ketbra{\Psi(T)}_{RQ}) &\leq& \sqrt{\Pur\left(\frac{S}{d^{\alpha}},\frac{T}{d^{\alpha}}\right)} \\
&\leq& \sqrt{\delta},
\end{eqnarray*} 
we obtain a randomness assisted protocol that makes an error of $10\eps + |\net|\delta' + \sqrt{\delta}$ for all $S$ (using the triangle inequality for purified distance, Fact \ref{fact:trianglepurified}). 

Letting $\delta' \rightarrow 0$ and then $\delta \rightarrow 0$, we obtain a randomness assisted protocol that makes an error of at most $11\eps$ for all $S$. The number of qubits communicated in the protocol is 
$$\frac{1+\alpha}{2}\log d + 4\log\frac{1}{\eps}$$
 and the number of ebits gained in the protocol is 
 $$\frac{1-\alpha}{2}\log d- 4\log\frac{1}{\eps}.$$

This completes the proof. 
\end{proof}

We use  above results to prove the following theorem.
\begin{theorem} \label{thm:worstAlpha-bitun}
Fix a quantum channel $\cN_{J\to K}$ and an $\alpha\in (0,1)$.  Let $\psi_{JJ'}$ be an arbitrary quantum state and define 
\begin{eqnarray*}
W &=& \frac{1}{2}I(J':K)_{\cN_{J\to K}(\psi_{JJ'})}, \\
V &=& \frac{1}{2}I(J':L)_{\cN^c_{J\to L}(\psi_{JJ'})}, \\
Y &=& \max(- \condent{J}{K}_{\cN_{J\to K}(\psi_{JJ'})}, 0).
\end{eqnarray*} For every $\eps > 0$, there exist $n, d$ large enough such that there exists a $(\log d, 4\eps, \alpha, n)$- entanglement unassisted protocol such that $n\log d$ divided by the number of channel use (or number of $\alpha$-bits transmitted per channel use) is equal to 
\vspace{1mm}
\begin{itemize}
\item $\frac{2}{1+\alpha}W,$ if $\frac{W}{V}\geq \frac{1+\alpha}{1-\alpha}$ (for this case, $\frac{2}{1+\alpha}W \leq \frac{Y}{\alpha}$) or \vspace{2mm}
\item $\frac{Y}{\alpha},$ if $\frac{W}{V}< \frac{1+\alpha}{1-\alpha}$ (for this case, $\frac{2}{1+\alpha}W \geq \frac{Y}{\alpha}$). \end{itemize} \vspace{1mm}
Thus the $\alpha$-bit capacity is equal to 
$$\min \left\{\frac{2}{1+\alpha}W, \frac{Y}{\alpha}\right\}.$$

\end{theorem}

\begin{proof}
The proof follows in a manner similar to the proof of Theorem \ref{unassistedav}. From Proposition \ref{stateredisttheo}, there exists a protocol for quantum state transfer of $\Psi(S)$ for all $S$, which makes an error of $11\eps$. Its quantum communication cost is 
$$\frac{1+\alpha}{2}\log d + 4\log\frac{1}{\eps},$$ and the number of ebits gained is $$\frac{1-\alpha}{2}\log d - 4\log\frac{1}{\eps}.$$ Using this protocol in Theorem \ref{unassistedav}, instead of the average case protocol that works for the uniform distribution over $S$, we obtain the result. 
\end{proof}

\subsection{Classical subset as promise with entanglement assistance}
\label{sec:classicalent}

There are two parties Alice  and Bob . Fix a set $S$ of size $|S|=2^{\alpha R}$. Alice  wants to communicate a classical message $M$ chosen uniformly from $[1:2^R]$ to Bob  over a quantum channel such that Bob  is able to decode the correct message with probability at least $1-\eps$ , for all message $m \in S$. Further, Alice is unaware of $S$, except for the value of $\alpha$. To accomplish this task Alice  and Bob  also share entanglement between them.  Let the input to Alice  be given in a register $M$. We now make the following definition:
\begin{definition}
Let $\ket{\theta}_{E_AE_B}$ be the state shared between Alice  and Bob . A $(R, \eps, \alpha )$-entanglement assisted code for classical communication over the quantum channel $\cN_{ A \to B}$ consists of 
\begin{itemize}
\item An encoding operation $\cE: ME_A \rightarrow A $ for Alice that does not depend on $S$.  
\item A decoding operation $\cD : B E_B\rightarrow M'$ for Bob, with $M'\equiv M$ being the output register such that for all $m$,
\beq
\Pr(M'\neq M\mid M \in S) \leq \eps.\nonumber
\enq
\end{itemize}
\end{definition}

\subsection*{A near-optimal achievability protocol}
Our achievability result will be based on a protocol for classical state redistribution, first obtained in \cite{BravermanRao11} for expected communication and made explicit for the worst case communication in \cite{AnshuJW17classical}. 
\begin{theorem}[\cite{BravermanRao11}, \cite{AnshuJW17classical}, Theorem 1]
\label{theo:BRraoalpha}
Fix an integer $R>0$ and $\delta \in (0,1)$. Let $MS$ be a joint random variable such that $S$ is distributed uniformly over all subsets of $[1: 2^R]$ of size $2^{\alpha R}$ and $M$ is uniformly distributed over elements of $S$. Let $M_1$ be a copy of $M$, that is $M_1M$ are perfectly correlated. Alice received a sample from $M$ and Bob receives a sample $s$ from $S$. There exists a randomness assisted one-way protocol in which Bob outputs a random variable $M'$ such that for all $s$, $$\|(MM' \mid S=s) - (MM_1 \mid S=s)\|_1 \leq \delta.$$ The number of bits communicated from Alice to Bob is 
$$\alpha R + 2\log\frac{1}{\delta}.$$ 
\end{theorem}

We show the following result. 
\begin{theorem}
\label{theo:achieveptop}
Let $\cN_{A \to B }$ be the quantum channel and let $\eps,\delta \in (0,1)$. Let $A'\equiv A$ be a purifying register. Then, for any $R$ satisfying 
\begin{multline}
R \leq  \frac{1}{\alpha}\max _{\ketbra{\psi}_{A A'}}\dmineps{\cN_{A \to B } (\ketbra{\psi}_{AA'})}{\cN_{A\to B}(\psi_{A}) \otimes \psi_{A'}}{\eps}\\ - \frac{4}{\alpha}\log \frac{2}{\delta},
\end{multline}
there exists a $(R, 2\eps + 3\delta, \alpha)$ entanglement assisted code for classical communication over the quantum channel $\cN_{A \to B }.$

Furthermore, for any $(R, \eps, \alpha)$ entanglement assisted code for classical communication over the quantum channel $\cN_{A\to B}$, it holds that 
$$R\leq  \frac{1}{\alpha}\max _{\ketbra{\psi}_{A A'}}\dmineps{\cN_{A \to B } (\ketbra{\psi}_{AA'})}{\cN_{A\to B}(\psi_{A}) \otimes \psi_{A'}}{\eps}.$$ 
\end{theorem}

\begin{proof}
The achievability proof is along the lines of the proof of Theorem \ref{theo:qachieve}, where we consider the task of classical state redistribution under a noisy channel. Combining Theorem \ref{theo:BRraoalpha} with the entanglement assisted protocol for communication over the channel $\cN_{A \to B}$ given in \cite[Theorem 1]{AnshuJW17CC}, we find that the largest possible value of $R$ which can be achieved with one use of the channel is 
\begin{multline*}
\alpha R + 2\log\frac{2}{\delta} \leq \\ \max _{\ketbra{\psi}_{A A'}}\dmineps{\cN_{A \to B } (\ketbra{\psi}_{AA'})}{\cN_{A\to B}(\psi_{A}) \otimes \psi_{A'}}{\eps} - 2\log\delta.
\end{multline*}

For the converse proof, we use the result given in \cite[Theorem 18]{MatthewsW14}. It is shown that for any $(\alpha\cdot R, \eps, 1)$ entanglement assisted code for the quantum channel $\cN_{A\to B}$, we have that 
 $$\alpha \cdot R\leq \max _{\ketbra{\psi}_{A A'}}\dmineps{\cN_{A \to B } (\ketbra{\psi}_{AA'})}{\cN_{A\to B}(\psi_{A}) \otimes \psi_{A'}}{\eps}.$$
Since having a $(R, \eps, \alpha)$ entanglement assisted code is a stronger requirement than having an $(\alpha\cdot R, \eps, 1)$ entanglement assisted code (as in the latter case, Alice knows the subset $S$), the upper bound follows. 
\end{proof}

\subsection{Classical subset as promise with randomness assistance}

\label{sec:classical}
There are two parties Alice  and Bob . Fix a set $S$ of size $|S|=2^{\alpha R}$. Alice  wants to communicate a classical message $M$ chosen uniformly from $[1:2^R]$ to Bob  over a quantum channel such that Bob  is able to decode the correct message with probability at least $1-\eps$ , for all message $m \in S$. Further, Alice is unaware of $S$, except for the value of $\alpha$. To accomplish this task Alice  and Bob  also share randomness between them.  Let the input to Alice  be given in a register $M$. We now make the following definition:
\begin{definition}
Let $\theta_{E_AE_B}$ be the shared randomness between Alice  and Bob. A $(R, \eps, \alpha )$-randomness assisted code for classical communication over the quantum channel $\cN_{ A \to B}$ consists of 
\begin{itemize}
\item An encoding operation $\cE: ME_A \rightarrow A $ for Alice that does not depend on $S$.  
\item A decoding operation $\cD : B E_B\rightarrow M'$ for Bob, with $M'\equiv M$ being the output register such that for all $m$,
\beq
\Pr(M'\neq M\mid M\in S) \leq \eps. \nonumber
\enq
\end{itemize}
\end{definition}

Along the lines similar to Theorem \ref{theo:achieveptop}, we obtain the following result. 
\begin{theorem}
\label{theo:achieverandptop}
Let $\cN_{A \to B }$ be the quantum channel and let $\eps,\delta \in (0,1)$. Let $\psi_{AA'}$ be a classical-quantum state where register $A'$ is classical. Then, for any $R$ satisfying 
\beq
R \leq  \frac{1}{\alpha}\max _{\psi_{A A'}}\dmineps{\cN_{A \to B } (\psi_{AA'})}{\cN_{A\to B}(\psi_{A}) \otimes \psi_{A'}}{\eps}- \frac{4}{\alpha}\log \frac{2}{\delta},
\enq
there exists a $(R, 2\eps + 2\delta, \alpha)$ randomness assisted code for classical communication over the quantum channel $\cN_{A \to B }.$
\end{theorem}

\section{Quantum state redistribution as a resource}
\label{sec:qsrresource}

In the previous sections, we considered simulating the task of quantum state redistribution with classical communication and entanglement assistance. In this section, we will show that quantum state redistribution can itself be used to simulate classical communication, if viewed as a resource. For this, we introduce the following definition.

\begin{definition}[Entanglement assisted simulation by Quantum State Redistribution (QSR)]
\label{def:qsrsim}
Fix an $\eps \in (0,1)$, a quantum state $\Psi_{RABC}$ and a $(q, \eps)$- quantum state redistribution protocol $\cP$ for $\Psi_{RABC}$. A $(N, \eps)$-QSR simulation consists of 
\begin{itemize}
\item An entangled quantum state $\ket{\omega}_{S_AS_B}$ shared between Alice $(E_A)$ and Bob $(E_B)$. 
\item For every $m\in [1:2^N]$ given in a register $M$, Alice chooses registers $A(m)C(m)$ (from her register $S_A$) holding the quantum state $\Psi_{A(m)C(m)}$ and runs the protocol $\cP$ with register $Q(m)$ obtained at Bob. 
\item A decoding map $\cD: \cL(S_BQ(m))\rightarrow \cL(M')$ such that $$\Pr[M'\neq m\mid M=m] \leq \eps.$$
\end{itemize} 
\end{definition}

Our one-shot simulation result is as follows. 

\begin{proposition}
\label{prop:qsrsim}
Fix $\eps, \delta\in (0,1)$ and a quantum state $\Psi_{RABC}$. Let $\cP$ be a $(q, \eta)$-quantum state redistribution protocol for $\Psi_{RABC}$. Then for any $N$ satisfying
\begin{multline*}N \leq  \dmineps{\Psi_{RBC}}{\Psi_{R}\otimes \Psi_{BC}}{\eps-2\sqrt{\delta}-\eta} \\ -  \dmaxeps{\Psi_{RB}}{\Psi_R\otimes \Psi_B}{\sqrt{\delta}} -  \log\frac{4\eps^3}{\delta^4},
\end{multline*}
there exists a $(N, \eps+2\delta)$-QSR simulation protocol that makes one use of $\cP$.
\end{proposition}
\begin{proof}
Fix the protocol $\cP$ which uses the shared entangled state $\ket{\theta}_{E_AE_B}$ and produces the message register $Q$ when run on $\Psi_{RABC}$. Let $\Omega_{RBQE_B}$ be the quantum state with Reference and Bob after Alice's message.  Let $\Psi'_{RB}$ be the quantum state achieving the optimum in the definition of $\dmaxeps{\Psi_{RB}}{\Psi_R\otimes \Psi_B}{\sqrt{\delta}}$ and let $\Psi'_{RABC}$ be its extension such that $\Psi'_{RABC}\in \ball{\sqrt{\delta}}{\Psi_{RABC}}$ (as guaranteed by Uhlmann's Theorem, Fact \ref{uhlmann}). Let $\Omega'_{RBQE_B}$ be the quantum state with Reference and Bob after Alice's message, if $\cP$ is run on $\Psi'_{RABC}$. Using Fact \ref{fact:monotonequantumoperation} and the fact that Bob has not performed any operation on $\Omega'_{RBQE_B}$, it holds that
\begin{equation}
\label{omegaprop} 
\Omega'_{RBQE_B}\in \ball{\sqrt{\delta}}{\Omega_{RBQE_B}}, \quad \Omega'_{RB}= \Psi'_{RB}.
 \end{equation}

\noindent {\bf Construction of the protocol:}  Define 
\begin{equation}
\label{defN'}
N' \defeq  \dmineps{\Omega'_{RBQE_B}}{\Omega'_{RB}\otimes \Omega'_{QE_B}}{\eps} - \log\frac{4\eps}{\delta^2}.
\end{equation}
We construct a $(N', \eps+2\delta)$-QSR simulation protocol as follows. We use the position-based decoding strategy introduced in \cite{AnshuJW17CC}. Alice and Bob share $2^{N'}$ copies of the quantum state $\Psi'_{RABC}$ in registers $R_1A_1B_1C_1, \ldots$ $R_{2^{N'}}A_{2^{N'}}B_{2^{N'}}C_{2^{N'}}$, where $R_iB_i$ belong to Bob and $A_iC_i$ belong to Alice (for all $i\in [1:2^{N'}]$). They additionally share the pure state $\ket{\theta}_{E_AE_B}$, where $E_A$ belongs to Alice and $E_B$ belongs to Bob. 

\vspace{0.1in}

\noindent{\bf Encoding:} To send the message $m\in [1:2^{N'}]$, Alice considers the registers $E_AA_mC_m$ and employs the protocol $\cP$ to produce the message register $Q$ with Bob. Observe that at this stage, the quantum state in registers $R_mB_mQE_B$ is $\Omega'_{R_mB_mQE_B}$ and the quantum state in registers $R_{m'}B_{m'}QE_B$ is $\Omega'_{R_{m'}B_{m'}}\otimes \Omega'_{QE_B}$.

\vspace{0.1in}

\noindent{\bf Decoding:} Let $\Pi_{RBQE_B}$ be the operator achieving the optimum in the definition of $\dmineps{\Omega'_{RBQE_B}}{\Omega'_{RB}\otimes \Omega'_{QE_B}}{\eps}$. Define 
$$\Lambda(m)\defeq \mathrm{I}_{R_1B_1} \otimes \mathrm{I}_{R_2B_2} \otimes \cdots \Pi_{R_mB_mQE_B} \otimes \cdots \otimes \mathrm{I}_{R_{2^{N'}}B_{2^{N'}}},$$ 
and 
$$\Gamma(m) \defeq \left(\sum_{ m' \in [1:2^{N'}]} \Lambda(m')\right)^{-\frac{1}{2}}\Lambda(m)\left(\sum_{m' \in [1:2^{N'}]} \Lambda(m')\right)^{-\frac{1}{2}}.$$
Bob applies the measurement $\{\Gamma(1), \ldots \Gamma(2^{N'}), \id - \sum_{m}\Gamma(m)\}$ to decode $m$.

\vspace{0.1in}

\noindent {\bf Error analysis:} Employing Hayashi-Nagaoka inequality (Fact \ref{haynag}), we have 
\begin{eqnarray*}
&&\Pr\{M'\neq m\} \\ && \leq (1+c) \Tr((\id-\Pi_{RBQE_B})\Omega'_{RBQE_B}) \\ 
&&\quad + \left(2+c+\frac{1}{c}\right)\cdot 2^{N'}\Tr(\Pi_{RBQE_B}\Omega'_{RB}\otimes \Omega'_{QE_B})\\
&& \leq (1+c)\eps + \frac{4}{c}2^{N'-\dmineps{\Omega'_{RBQE_B}}{\Omega'_{RB}\otimes \Omega'_{QE_B}}{\eps}}\\
&& \leq \eps+2\delta,
\end{eqnarray*}
where we choose $c = \frac{\delta}{\eps}$. 

\vspace{0.1in}

\noindent{\bf Lower bounding $N'$:} Now, we lower bound $N'$ in terms of $N$ (as defined in the statement of the theorem). Let $\Gamma$ be the operator achieving the optimum in the definition of $$K \defeq \dmineps{\Omega_{RBQE_B}}{\Psi_{R}\otimes \Psi_B\otimes \Omega'_{QE_B}}{\eps-\sqrt{\delta}}.$$
From Fact \ref{fact:err} and Equation \ref{omegaprop}, we conclude 
$$\Tr(\Gamma\Omega'_{RBQE_B}) \geq 1- \eps.$$ Further, 
\begin{eqnarray*}
2^{-K} &=& \Tr(\Gamma \Psi_{R}\otimes \Psi_B\otimes \Omega'_{QE_B}) \\ 
 &\geq& 2^{-\dmaxeps{\Psi_{RB}}{\Psi_R\otimes \Psi_B}{\sqrt{\delta}}}\Tr(\Gamma \Psi'_{RB}\otimes \Omega'_{QE_B})\\
&=& 2^{-\dmaxeps{\Psi_{RB}}{\Psi_R\otimes \Psi_B}{\sqrt{\delta}}}\Tr(\Gamma \Omega'_{RB}\otimes \Omega'_{QE_B})\\
&\geq & 2^{-\dmaxeps{\Psi_{RB}}{\Psi_R\otimes \Psi_B}{\sqrt{\delta}}}\cdot 2^{- \dmineps{\Omega'_{RBQE_B}}{\Omega'_{RB}\otimes \Omega'_{QE_B}}{\eps}},
\end{eqnarray*}
where we have used the definition of $\Psi'_{RB}$ in the first inequality and Equation \ref{omegaprop} in the second equality. The last inequality follows from the definition of $\dmineps{\Omega'_{RBQE_B}}{\Omega'_{RB}\otimes \Omega'_{QE_B}}{\eps}$. Thus, 
\begin{multline}
\label{eq:chainrule1}
\dmineps{\Omega'_{RBQE_B}}{\Omega'_{RB}\otimes \Omega'_{QE_B}}{\eps} + \dmaxeps{\Psi_{RB}}{\Psi_R\otimes \Psi_B}{\sqrt{\delta}} \\ \geq \dmineps{\Omega_{RBQE_B}}{\Psi_{R}\otimes \Psi_B\otimes \Omega'_{QE_B}}{\eps-\sqrt{\delta}}.
\end{multline}
For the protocol $\cP$, let $W$ be Bob's unitary after Alice's message, as given in Definition \ref{def:qsr}. Define 
$$\Theta_{RBCT_B} \defeq W\Omega_{RBQE_B}W^{\dagger}.$$ It holds that $\Theta_{RBC}\in  \ball{\eta}{\Psi_{RBC}}$. Thus, we have
\begin{eqnarray*}
&&\dmineps{\Omega_{RBQE_B}}{\Psi_{R}\otimes \Psi_B\otimes \Omega'_{QE_B}}{\eps-\sqrt{\delta}} \\
&=& \dmineps{\Theta_{RBCT_B}}{\Psi_{R}\otimes W(\Psi_B\otimes \Omega'_{QE_B})W^{\dagger}}{\eps-\sqrt{\delta}} \\
&\geq& \inf_{\sigma_{BC}} \dmineps{\Theta_{RBC}}{\Psi_{R}\otimes \sigma_{BC}}{\eps-\sqrt{\delta}}\\
&\geq& \inf_{\sigma_{BC}} \dmineps{\Psi_{RBC}}{\Psi_{R}\otimes \sigma_{BC}}{\eps-\sqrt{\delta}-\eta}\\
&\geq& \dmineps{\Psi_{RBC}}{\Psi_{R}\otimes \Psi_{BC}}{\eps-2\sqrt{\delta}-\eta} - \log\frac{\eps^2}{\delta},
\end{eqnarray*}
where the second last inequality uses Fact \ref{fact:smoothdh} and the last inequality uses Fact \ref{fact:Ihdhsame}. Combining with Equation \ref{eq:chainrule1}, we obtain
\begin{multline*}\dmineps{\Omega'_{RBQE_B}}{\Omega'_{RB}\otimes \Omega'_{QE_B}}{\eps} \geq \\ \dmineps{\Psi_{RBC}}{\Psi_{R}\otimes \Psi_{BC}}{\eps-2\sqrt{\delta}-\eta}  -  \dmaxeps{\Psi_{RB}}{\Psi_R\otimes \Psi_B}{\sqrt{\delta}} \\ -  \log\frac{\eps^2}{\delta}.
\end{multline*}
From the definition of $N'$ in Equation \ref{defN'}, we conclude that 
\begin{multline*}N' \geq  \dmineps{\Psi_{RBC}}{\Psi_{R}\otimes \Psi_{BC}}{\eps-2\sqrt{\delta}-\eta} \\ -  \dmaxeps{\Psi_{RB}}{\Psi_R\otimes \Psi_B}{\sqrt{\delta}}  -  \log\frac{4\eps^3}{\delta^4}.
\end{multline*}
Thus, $N\leq N'$ and hence there exists a $(N, \eps+2\delta)$-QSR simulation protocol. 
\end{proof}

Now we are in a position to prove the main result of this section. 
\begin{theorem}
\label{theo:QSRsim}
Fix a quantum state $\Psi_{RABC}$. For every $\eps, \delta\in (0,1)$, there exists a $n$ large enough such that for any $(q, \eps)$- quantum state redistribution protocol for $\Psi^{\otimes n}_{RABC}$, there exists a $(n(N -\delta), 8\eps)$-QSR simulation for any $N$ satisfying
$$N \leq \condmutinf{R}{C}{B}_{\Psi}.$$
\end{theorem}
\begin{proof}
Let $n$ be an integer to be chosen later. From Proposition \ref{prop:qsrsim}, for any $(q, \eps)$-quantum state redistribution protocol for $\Psi^{\otimes n}_{RABC}$, there exists a $(M, 8\eps)$-QSR simulation protocol, if 
\begin{multline*}M \leq  \dmineps{\Psi^{\otimes n}_{RBC}}{\Psi^{\otimes n}_{R}\otimes \Psi^{\otimes n}_{BC}}{\eps}\\ -  \dmaxeps{\Psi^{\otimes n}_{RB}}{\Psi^{\otimes n}_R\otimes \Psi^{\otimes n}_B}{\eps}  -  10\log\frac{1}{\eps}.
\end{multline*} 
Using Fact \ref{dmaxequi}, we conclude that it suffices to have
$$M \leq  n\Biggl(\mutinf{R}{BC}_{\Psi} - \mutinf{R}{B}_{\Psi} - O\left(\sqrt{\frac{\log\frac{1}{\eps}}{n}}\right)\Biggr).$$ 
Choosing $n$ large enough such that $\delta \geq O\left(\sqrt{\frac{\log\frac{1}{\eps}}{n}}\right)$, and letting $N= \frac{M}{n}$, the proof concludes.  
\end{proof}

As its corollary, we obtain the entanglement assisted classical capacity of Alpha-bit, recovering the result shown in \cite{1706.09434}. We also use an argument derived from the subspace decoupling duality  \cite[Theorem 2]{1706.09434}.
\begin{corollary}
\label{cor:Alpha-bitsim}
Fix a protocol $\cP$ as given in Definition \ref{qcode}, where Bob is given a subspace $S$, Alice and Reference share a maximally entangled state $\ket{\Psi(S)}_{RQ}$ with support of $\Psi(S)_{Q}$ equal to $S$ and the protocol achieves the transfer of register $Q$ to Bob. The achievable rate of entanglement assisted classical protocol $\cP$ is at least $(1+\alpha)\log|Q|$.
\end{corollary}
\begin{proof}
In the protocol $\cP$, Alice and Bob also share an entangled state $\ket{\theta}_{E_AE_B}$. As argued in \cite[Theorem 5]{1706.09434}, using the subspace decoupling duality \cite[Theorem 2]{1706.09434}, the final state with Bob can be assumed to be close to $\ket{\Psi(S)}_{RQ'}\otimes \ket{\theta'}_{T_AT_B}$ (where $Q'\equiv Q$ is held with Bob), for some fixed state $\ket{\theta'}_{T_AT_B}$ independent of $S$.

Now, denoting the uniform distribution over the subspaces $S$ by $\mu(S)$, consider the quantum state 
$$\ket{\Psi}_{R_SRSQ}\defeq \sum_S\sqrt{\mu(S)}\ket{S, S}_{R_SS}\ket{\Psi(S)}_{RQ},$$ where $R, R_S$ is held by Reference, $Q$ is held by Alice and $S$ is held by Bob. Running the protocol $\cP$ on this quantum state leads to a final quantum state 
$$\sum_S\sqrt{\mu(S)}\ket{S, S}_{R_SS}\ket{\Psi(S)}_{RQ'}\otimes \ket{\theta'}_{T_AT_B},$$ where $Q', S$ are now held by  Bob. Thus, $\cP$ achieves the quantum state redistribution (more precisely, the quantum state merging) of $\ket{\Psi}_{R_SRSQ}$. From Theorem \ref{theo:QSRsim}, its entanglement assisted classical capacity is at least 
\begin{eqnarray*}
&& \condmutinf{RR_S}{Q}{S}_{\Psi} \\
&=& \mutinf{RR_S}{Q}_{\Psi} \\ 
&=& \mutinf{R_S}{Q}_{\Psi} + \condmutinf{R}{Q}{R_S}_{\Psi} \\ &=& \entr{\Psi_Q} - \condent{Q}{R_S}_{\Psi} + \condmutinf{R}{Q}{R_S}_{\Psi}\\
&=& \log|Q| - \alpha\log|Q| + 2\alpha\log|Q| \\
&=& (1+\alpha)\log|Q|.
\end{eqnarray*}
This completes the proof. 
\end{proof}

\section{Conclusion}

In this paper, we study the communication paradigm with one-sided promises that are available to the receiver, and characterize its theoretical communication capability in very general frameworks. Our results are obtained via a noisy version of quantum state redistribution, which might be of independent interest. In particular, we study two special cases of one-sided promises that are natural, namely, (i) the message set from which a message is chosen, and (ii) the description of the quantum subspace from which a quantum state is to be transmitted.  The latter was studied recently under the name of the ``Alpha-bit'' \cite{1706.09434}. As a result, we recover their asymptotic Alpha-bit capacities. 

An interesting observation drawn from these two cases is that the saving for quantum communication with a subspace promise is at most a factor of 2 than that without a promise. However, classical communication over a quantum channel with a subset promise can be achieved much more efficiently, given by $1/\alpha$, where $0<\alpha\leq 1$ provides information of the subset's size. 
%This provides, as far as we are aware, the first example of a separation between the communication resources when classical or quantum communication with one-sided promises is considered.  

\section*{Acknowledgement}
We thank Patrick Hayden and Geoffrey Penington for their helpful discussions on the Alpha-bit capacity. This work was done when A. A. was with the Centre for Quantum Technologies (CQT), Singapore. Research at CQT is supported by the Singapore Ministry of Education and the National Research Foundation also through the Tier 3 Grant ``Random numbers from quantum processes'' MOE2012-T3-1-009. RJ is also supported by the ``NRF2017-NRF-ANR004 VanQuTe'' grant and by the ``VAJRA Faculty Scheme'' of the Science and Engineering Board
(SERB), Department of Science and Technology (DST), Government of India.
 M.-H. H. is supported in part by an ARC Future Fellowship under Grant FT140100574 and in part by U.S. the Army Research Office for Basic Scientific Research under Grant W911NF-17-1-0401.

%\bibliographystyle{ieeetr}
%\bibliography{References}

\end{document}